\documentclass[11pt]{article}
\usepackage{graphicx}
\usepackage{amsmath}
\usepackage{float}
\usepackage{color}

\def\RR{{\rm I\kern-.17em R}}
\usepackage{amsmath}
\usepackage{amssymb}
\usepackage{amsfonts}
\def\RR{{\rm I\kern-.17em R}}
\usepackage{color}
\usepackage{graphicx}
\usepackage{float}
\usepackage{psfrag}
\usepackage{graphics}
\newtheorem{lemma}{Lemma}
\newtheorem{proposition}{Proposition}

\newcommand{\be}{\begin{eqnarray}}

\newcommand{\ee}{\end{eqnarray}}

\begin{document}
\title{Geodesics dynamics in the Linet-Tian spacetime with $\Lambda>0$}
\author{Irene Brito$^\flat$, M. F. A. Da Silva$^\star$, Filipe C. Mena$^{\flat\star}$ and N. O. Santos$^{\flat\flat\star}$\\\\
$^\flat$Centro de Matem\'atica, Universidade do Minho, 4710-057 Braga, Portugal\\
$^{\star}$Departamento de F\'{i}sica Te\'orica, Instituto de F\'isica, Universidade do Estado do Rio de Janeiro,\\ Rua S\~ao Francisco Xavier 524, Maracan\~a, 20550-900, Rio de Janeiro, Brasil\\
$^{\flat\flat}$School of Mathematical Sciences, Queen Mary, University of London, London E1 4NS, U.K.\\
$^{\flat\flat}$Sorbonne Universit\'es, UPMC Universit\'e Paris 06, LERMA, UMRS8112 du CNRS, \\
Observatoire de Paris-Meudon, 5, Place Jules Janssen, F-92195 Meudon cedex, France}

\maketitle
\begin{abstract}
We analyse the geodesics' dynamics in cylindrically symmetric vacuum spacetimes with $\Lambda>0$ and compare it to the $\Lambda= 0$ and $\Lambda<0$ cases. When $\Lambda>0$ there are two singularities in the metric which brings new qualitative features to the dynamics.

We find that $\Lambda=0$ planar timelike confined geodesics are unstable against the introduction of a sufficiently large $\Lambda$, in the sense that the bounded orbits become unbounded. In turn, any non-planar radially bounded geodesics are stable for any positive $\Lambda$.

We construct  global non-singular static vacuum spacetimes in cylindrical symmetry with $\Lambda>0$ by matching the Linet-Tian metric with two appropriate sources.
\end{abstract}

\section{Introduction}

Globally cylindrically symmetric solutions of the Einstein field equations (EFEs), at first sight, may not seem to be physically relevant since they impose infinitely long sources. Nonetheless, under controlled circumstances, they can provide fairly accurate descriptions of different physical phenomena.

In Newtonian theory, the gravitational field produced by an infinite cylindrical source has a potential $U=2\sigma\ln\rho$, in cylindrical polar coordinates $(\rho,z,\phi)$ such that the axis lies at $\rho=0$,  where $\sigma$ is the mass per unit length of the source. Working out the corresponding field in Einstein's theory of gravitation, which has been done by Levi-Civita \cite{Levi} in 1919, one obtains a similar potential for the Newtonian limit with a parameter $\sigma$. However, as demonstrated by Marder \cite{Marder} in 1958, a second independent parameter emerges, which has been interpreted as a topological defect \cite{Dowker,Bonnorx,Bonnor}. Unexpectedly, the parameter $\sigma$ is the most difficult and elusive to be interpreted and there is a long list of articles dedicated to unveil its meaning (see e.g. \cite{Herrera,Griffiths}). In spite of that, the Newtonian limit in cylindrical models agrees well with observations \cite{Fujimoto, Hockney,Song}.

Due to the non-triviality of the topological parameters which arise naturally in this symmetry, their study may help to reveal some internal features of the theory of General Relativity, which are not present in spherical symmetry, such as the gravitational analogue of the Aharonov-Bohm effect, where particles constrained to move in a region where the Riemann tensor vanishes can, however, have physical effects arising from non-zero curvature in a region from which they are excluded \cite{Dowker, FV, HoMorgan, Stachel, Bezerra}.

General relativistic cylindrical systems have also been used to improve our compreehention about many other astrophysical situations such as: the dragging of spacetime due to rotation which produces new topological defects \cite{Herrera1,MacCallum,Mashhoon}; exact models of rotation \cite{Mashhoon}; topological cosmic strings that might have been formed in the early stages of the Universe \cite{Vilenkin,Hindmarsh,Silva}; models for extragalactic jets \cite{Opher,Herrera3,Gariel}; gravitational collapse \cite{Apostolatos,Pereira} and its eventual gravitational radiation \cite{Prisco,Herrera2,MacCallum}; translating fluids with cylindrical symmetry \cite{Griffiths10} have been used to study beams of light that might be produced by stars \cite{Bonnor10,Bonnor20}.

Supernovae observations \cite{supernovae} have motivated the inclusion of a positive cosmological constant $\Lambda$ in the EFEs. Physically, this constant is often  interpreted as the vacuum energy or some unknown dark energy (see e.g. \cite{Peebles}), which may exist in the universe as a whole, and therefore having an impact on the dynamics of local gravitational systems, such solar systems and galaxies (see e.g. \cite{galactic}).

Local gravitational systems can be modelled using symmetric exact solutions of the EFEs, such as the Schwarzschild solution in spherical symmetry or the Levi-Civita (LC) in cylindrical symmetry. The generalisations of these solutions to include $\Lambda$ are known as the Kottler and  Linet-Tian (LT) solutions, respectively.

In this paper, we will consider the LT solutions which, for $\Lambda > 0$, contain two curvature singularities and represent, at most,  the gravitational field in the region between two cylindrical sources.
The question of finding possible sources can be tackled using spacetime matching theory and finding the appropriate boundary hypersurfaces. Spacetime matchings have already been performed with LT as the exterior \cite{Brito,Brito2,GP,Zofka-Bicak} and we shall use these results in order to construct the first global non-singular solution involving LT with $\Lambda>0$.

In particular, we construct a model with an interior anisotropic fluid source, an intermediate Linet-Tian spacetime and an exterior containing $\Lambda$-dust matter. Physically, this solution can model an intermediate $\Lambda$-vacuum medium between an astrophysical local object and a background where the object is embedded.


In this paper, we will mostly study geodesics and these can be used to model jets on cylindrically symmetric regions.  
In nature, jets are observed at scales ranging from sub-parsecs up to hundreds of kiloparsecs \cite{Marscher}.
The most often invoked process of production of these jets is the Blandford-Znajek
\cite{BlandfordZnajek}
in the framework of magnetohydrodynamics.  More recently, a model to explain the origin and structure of jets, has been put forward based on a purely general relativistic origin
\cite{Gariel, Freitas, Gariel1, Gariel2}.
That model considers the jet originating from a Penrose process
\cite{Penrose}
in the ergosphere of a Kerr black hole. The ejected particles undergo the effect of strong repulsive gravitational fields, produced by the Kerr black hole, and follow geodesics that collimate them.

The solution presented here can represent a  very simple and approximated model to that region, e.g. on an extra galactic scale, since a jet typically emerges outwards from the centre of the galactic disc, and perpendicular to it. Then, we can have the material ejected from some rotating massive and compact central object, or another alternative mechanism, forming a cylindrical flow, confined by the layer of a  LT vacuum. In this context, the possible relevance of our setting is not to show the origin of jets, but to show whether the cylindrical symmetry can furnish a purely gravitational contribution to explain the collimation of the (extragalactic) jets.

Since geodesics describe the orbits of light rays and material particles, an important aspect, from both the physical and mathematical points of view, is the influence of the positive cosmological constant on the stability of the geodesics orbits in vacuum cylindrically symmetric systems with $\Lambda=0$.
The LT solution is the unique static $\Lambda$-vacuum spacetime in cylindrical symmetry and reduces to the exact vacuum case for $\Lambda=0$, thus providing an appropriate framework to study stability problems.

We shall then investigate the impact of arbitrarily small values of $\Lambda$ on the kinematics and dynamics of the geodesic motion, which we then compare with the $\Lambda=0$ case. In this context, whenever we talk about the geodesics' stability we shall mean the stability, in the sense of bounded-unbounded orbits, with respect to changes in $\Lambda$. In this aspect, this is a twin paper to \cite{BDMS-2014}, where the $\Lambda<0$ solution, which has only one singularity, has been analysed. Contrary to that case, which had been previously studied \cite{Banerjee}, the case $\Lambda>0$ has never been analysed before.
As we shall see ahead, there will be important differences in the geodesics' stability with respect to the $\Lambda< 0$ case, not only because of the sign change in $\Lambda$, which naturally affects the dynamics, but also due to the existence of the second singularity.

The paper is organised as follows: In Section 2, we revise the LT spacetimes and their geodesics equations and, in Section 3, we construct a globally non-singular spacetime by matching the LT solution to appropriate sources. Section 4 is devoted to the particular case of circular geodesics, while Section 5 concentrates on the dynamics along the $z$ direction. Section 6 is the longest and contains the most general treatment about geodesics along the radial direction. We split our analysis into planar and non-planar as well as null and non-null geodesics. In most cases, we illustrate our analytical results with plots of numerical examples. Throughout we use units such that $c=G=1$.
\section{The LT spacetime in brief}
We recall that the LT metric with $\Lambda>0$ is given by \cite{L,T}
\begin{equation}
ds^2=-fdt^2+d\rho^2+gdz^2+ld\phi^2, \label{g1}
\end{equation}
where
\begin{eqnarray}
f=a^2 Q^{2/3}P^{-2(1-8\sigma+4\sigma^2)/3\Sigma}, \label{g2}\\
g=b^2Q^{2/3}P^{-2(1+4\sigma-8\sigma^2)/3\Sigma}, \label{g3}\\
l=c^2Q^{2/3}P^{4(1-2\sigma-2\sigma^2)/3\Sigma}, \label{g4}
\end{eqnarray}
with
\begin{equation}
P=\frac{2}{\sqrt {3\Lambda}}\tan R, \;\; Q=\frac{1}{\sqrt {3\Lambda}}\sin(2R), \;\; R=\frac{\sqrt {3\Lambda}}{2}\rho \label{1}
\end{equation}
and $t$, $\rho$, $z$ and $\phi$ are cylindrical coordinates, $\Sigma=1-2\sigma+4\sigma^2$, the constant $\sigma$ is related to the mass per unit length and $b$ and $c$ are constants related to the angle defects \cite{Bonnor,Bonnor1,Wang}. For $1/2<\sigma<\infty$, the spacetime description is similar to the case $0\leq\sigma\leq 1/2$, by redefining $\sigma$  \cite{GP,Herrera3},  therefore we assume $0\leq\sigma\leq 1/2$. For $\Lambda=0$, the metric reduces to the Levi-Civita metric \cite{Levi}, in which case $P=Q=\rho$.

The coordinate $\rho$ lies in the range $\rho\in (0, \pi/\sqrt{3\Lambda})$, corresponding to $R\in(0,\pi/2)$, and there are singularities at $\rho=0$ and $\rho=\pi/\sqrt{3\Lambda}$, where the Kretschmann scalar diverges.
As in the $\Lambda\le 0$ case, one can check that there are no trapped cylinders, by using the expression (7) of \cite{BDMS-2014}.
The geodesics equations for LT can also be found in \cite{BDMS-2014} and imply
\begin{eqnarray}
{\dot t}=\frac{E}{f}, \label{gg5}\\
{\dot\rho}^2=\frac{E^2}{f}-\epsilon-\frac{P_z^2}{g}-\frac{L_z^2}{l}, \label{gg6}\\
{\dot z}=\frac{P_z}{g}, \label{gg7}\\
{\dot\phi}=\frac{L_z}{l}, \label{gg8}
\end{eqnarray}
where the dot stands for differentiation with respect to an affine parameter $\lambda$. In turn, $\epsilon=0$, $1$ or $-1$ if the geodesics are, respectively, null, timelike or spacelike, and the constants $E$, $P_z$ and $L_z$ represent, respectively, the total energy of the test particle, its momentum along the $z$ axis and its angular momentum about the $z$ axis.

\section{Two sources for LT and a global non-singular solution}

Our goal here is to construct a global non-singular solution by substituting the two singularities of the LT spacetime with appropriate sources. The procedure will involve matching the LT metric to an interior metric  at some constant value  of $\rho$, say $\rho_2$, and to an exterior metric at some $\rho_1$, with $\rho_1>\rho_2$.

It is already known that LT matches with the Einstein static universe \cite{GP} and with a conformally flat source \cite{Brito}. However, none of these solutions can, alone, provide the two sources of LT. We shall then consider the former as the exterior to be matched at $\rho_1$ and the latter as the interior matched at $\rho_2$. In particular, we shall have to ensure the consistency of the model since the interior and exterior will also be related through the (local) matching conditions at both ends.

The matching of any two $\mathcal{C}^2$ spacetimes $(M^\pm,g^\pm)$ with non-null boundaries $\Sigma^\pm$ requires the identification of the boundaries, i.e.  embeddings $\Phi_\pm: \Sigma\to M^\pm$ with $\Phi_{\pm}(\Sigma)=\Sigma^\pm$, where $\Sigma$ is an abstract copy of either boundary. We denote coordinates in $\Sigma$ by $\xi^\alpha$, $\alpha=1,2,3$, orthonormal tangent vectors to $\Sigma^\pm$ by $ e_\alpha^{\pm i}$ and normal vectors by $n_\pm^ i$. The first and
second fundamental forms at $\Sigma^\pm$ are $q^\pm_{\alpha \beta}=e_\alpha^{\pm i} e_\beta^{\pm j} g^\pm_{i j}$ and $H^\pm_{\alpha \beta}=-n^\pm_i e_\alpha^{\pm j} \nabla^\pm_ j e_\beta^{\pm i}$.
The necessary and sufficient conditions for the matching (in the absence of shells) are:
\begin{equation}
\label{}
q_{\alpha \beta}^-=q_{\alpha \beta}^+,~~~~H_{\alpha \beta}^-=H_{\alpha \beta}^+.
\end{equation}
Applying these to our case we shall consider the matching between LT with the Einstein static universe (the exterior) at $\rho_1$ and with a conformally flat source (the interior) at $\rho_2$. We shall then have to ensure the consistency of the model since the interior and exterior will also be related through the local matching conditions at both ends.
\subsection{The exterior}
The metric of the Einstein static universe can be written in cylindrical symmetry as \cite{GP}
\begin{equation}
\label{EUniv}
ds^2= -A_{1}^{2}dt^2+\frac{B_{1}^{2}}{\Lambda}\cos^{2}[\sqrt{\Lambda}(R-R_0)]dz^2 +\frac{C_{1}^{2}}{\Lambda}\sin^{2}[\sqrt{\Lambda}(R-R_0)]d\phi^2+dR^2,
\end{equation}
where $\phi\in[0,2\pi)$ and $A_1,$ $B_1,$ $C_1,$ $R_0$ are non-zero constants.
The matching of \eqref{EUniv} with \eqref{g1}, across a cylindrical surface $S_1$ as measured by $R=R_1$, which can be identified with $\rho_1$ implying the following conditions from the equality of the second fundamental forms \cite{GP}:
\begin{equation}
\label{dm1}
\cos(\sqrt{3\Lambda}\rho_1)\stackrel{S_1}{=}\frac{1-8\sigma+4\sigma^2}{\Sigma}.
\end{equation}
Given $\sigma$ and $\Lambda$, this equation determines an unique value for $\rho_1$,
 and
 \begin{equation}
\label{m22}
\tan^2[\sqrt{\Lambda}(\rho_1-R_0)]\stackrel{S_1}{=}\frac{4\sigma(1-\sigma)}{1-4\sigma}
\end{equation}
defines an unique positive value for the parameter $R_0$.
The values of the constants $A_1$, $B_1$ and $C_1$ are obtained from the equality of the first fundamental forms as \cite{GP}
\begin{align}
A_1\stackrel{S_1}{=}aQ(\rho_1)^{1/3}P(\rho_1)^{-(1-8\sigma+4\sigma^2)/3\Sigma},\label{m1}\\
B_1\stackrel{S_1}{=}b\sqrt{\Lambda}\,\frac{Q(\rho_1)^{1/3}P(\rho_1)^{-(1+4\sigma-8\sigma^2)/3\Sigma}}
{\cos[\sqrt{\Lambda}(R_1-R_0)]},\label{m2}\\
C_1\stackrel{S_1}{=}c\sqrt{\Lambda}\,\frac{Q(\rho_1)^{1/3}P(\rho_1)^{2(1-2\sigma-2\sigma^2)/3\Sigma}}
{\sin[\sqrt{\Lambda}(R_1-R_0)]}.\label{m3}
\end{align}
So, in this case, given the interior parameters $\Lambda, \sigma, a, b$ and $c$, the matching conditions provide five independent equations which determine uniquely the four parameters $A_1,$ $B_1,$ $C_1,$ $R_0$ of \eqref{EUniv} and the location $R_1$ of the boundary.

\subsection{The interior}

The conformally flat cylindrically symmetric interior source will be \cite{Brito}
\begin{equation}
\label{conf-flat}
ds^2=\frac{1}{(a_4[\cosh(a_2 r)-1]+1)^2}\left[-\cosh^2(a_2 r)dt^2 +dr^2+dz^2+\frac {\sinh^2(a_2 r)}{a^2_2} d\phi^2\right],
\end{equation}
where $a_2\neq0$ and $a_4\neq 0$ are constants.
The matching of \eqref{conf-flat} with \eqref{g1}, across a cylindrical surface $S_2$ as measured by $r=r_2$,
was performed in \cite{Brito}, where it was also shown that \eqref{conf-flat} has a regular centre.
 The matching conditions from the equality of the second fundamental forms are the following \cite{GP}:
 \begin{eqnarray}
\label{press} a_2^2 \stackrel{S_2}{=}\Lambda \left[2a_4-1-2a_4(a_4
-1)\frac{4\sigma(1-\sigma)}{\sqrt{(1-4\sigma)(1-4\sigma^2)}}\right]^{-1},
\end{eqnarray}
which determines $0<\sigma<1/4$, given the interior data $a_2$, $1/2\leq a_4 \leq 1$ and $\Lambda>0$, and
\begin{eqnarray}
\label{dm2} \sin^2 \left(\frac{\sqrt{3\Lambda}}{2}\rho_2 \right) \stackrel{S_2}{=} \frac{3
\sigma}{\Sigma}\left[1+ \frac{2(a_4 -1) (1-\sigma)
\sqrt{1-4\sigma}}{a_4 \sqrt{1-4\sigma^2}- (a_4
-1)\sqrt{1-4\sigma}}\right],
\end{eqnarray}
which determines $\rho_2$, together with
\begin{eqnarray}
\label{sig6} \sinh^2(a_2
r_2)\stackrel{S_2}{=}\frac{4\sigma(1-\sigma)}{1-4\sigma},
\end{eqnarray}
which fixes $r_2$. In turn, the parameters $a$, $b$ and $c$ are obtained from equating the first fundamental forms as \cite{Brito}
\begin{eqnarray}
\frac{\cosh(a_2 r_2)}{a_4[\cosh(a_2 r_2)-1]+1}\stackrel{S_2}{=}a\,Q(\rho_2)^{1/3}P(\rho_2)^{-(1-8\sigma+4\sigma^2)/3\Sigma}, \label{Mc31}\\
\frac{1}{a_4[\cosh(a_2 r_2)-1]+1}\stackrel{S_2}{=}b\,Q(\rho_2)^{1/3}P(\rho_2)^{-(1+4\sigma-8\sigma^2)/3\Sigma}, \label{Mc32}\\
\frac{\sinh(a_2 r_2)}{a_2\{a_4[\cosh(a_2 r_2)-1]+1\}}\stackrel{S_2}{=}c\,Q(\rho_2)^{1/3}P(\rho_2)^{2(1-2\sigma-2\sigma^2)/3\Sigma}. \label{Mc33}
\end{eqnarray}
So, given the interior parameters $a_2, a_4$ and $\Lambda$, the matching conditions provide six independent equations which determine uniquely the parameters $\sigma, a, b$ and $c$ of \eqref{g1} as well as the boundary, defined by the values of $\rho_2$ and $r_2$.

\subsection{Consistency and global solutions}
\begin{figure}[H]
\begin{center}
\includegraphics[width=8cm]{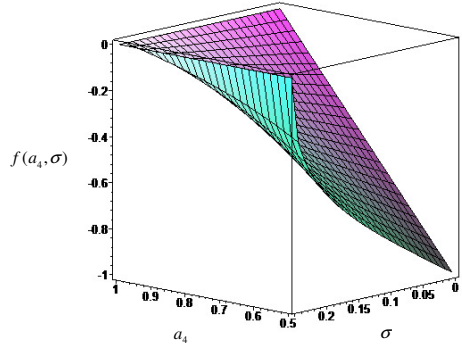} \\
\caption{Graph of $f(a_4, \sigma)$.}
\label{figa4sigma}
\end{center}
\end{figure}

In order to ensure the consistency of the double matching, one must ensure that $\rho_1 >\rho_2$, which, using (\ref{dm1}) and (\ref{dm2}) 
is equivalent to prove that
\begin{equation}
\frac{1}{2}\left(1-\frac{1-8\sigma+4\sigma^2}{\Sigma}\right)> \frac{3
\sigma}{\Sigma}\left[1+ \frac{2(a_4 -1) (1-\sigma)
\sqrt{1-4\sigma}}{a_4 \sqrt{1-4\sigma^2}- (a_4
-1)\sqrt{1-4\sigma}}\right].
\end{equation}
This condition can be rewritten as
\begin{equation}
\label{c1}
\frac{6\sigma}{\Sigma}> \frac{6
\sigma}{\Sigma}\left[1+ \frac{2(a_4 -1) (1-\sigma)
\sqrt{1-4\sigma}}{a_4 \sqrt{1-4\sigma^2}- (a_4
-1)\sqrt{1-4\sigma}}\right]
\end{equation}
and it can be checked that the value of the expression inside the brackets on the right hand side is always in the
interval $]0,1[$, for every $0<\sigma<1/4$ and $1/2\leq a_4 < 1$. If $a_4 =1,$ then $\rho_1 =\rho_2$ and the  LT spacetime can no longer represent the region between the two sources.
In fact, the distance between $\rho_1$ and $\rho_2$ is controlled by the function
\begin{equation}
\label{c2}
f(a_4,\sigma)=\frac{2(a_4 -1) (1-\sigma)
\sqrt{1-4\sigma}}{a_4 \sqrt{1-4\sigma^2}- (a_4
-1)\sqrt{1-4\sigma}},
\end{equation}
which satisfies
$-1 < f(a_4,\sigma)<0.$
The maximum distance between $\rho_1$ and $\rho_2$ is obtained when $f(a_4,\sigma)\rightarrow -1,$ which is achieved for $a_4 = 1/2$ and $\sigma\rightarrow 0$ (see also Figure \ref{figa4sigma}).

With this construction, the exterior parameters are related to the interior ones through the LT parameters.
Consistency relationships between the parameters can easily be obtained from the matching conditions.
The expressions for the parameters $a,b$ and $c$ obtained from \eqref{Mc31}-\eqref{Mc33} must be equal to those obtained from \eqref{m1}-\eqref{m3}, yielding
\begin{align}
A_1=\frac{\cosh(a_2 r_2)}{a_4 [\cosh(a_2 r_2)-1]+1}\left[\frac{Q(\rho_2)}{Q(\rho_1)}\right]^{-1/3}\left[\frac{P(\rho_2)}
{P(\rho_1)}\right]^{(1-8\sigma+4\sigma^2)/3\Sigma},\label{dm3}\\
\frac{B_1 \cos[\sqrt{\Lambda}(R_1 -R_0])}{\sqrt{\Lambda}}=\frac{1}
{a_4 [\cosh(a_2r_2)-1]+1}\left[\frac{Q(\rho_2)}{Q(\rho_1)}\right]^{-1/3}
\left[\frac{P(\rho_2)}{P(\rho_1)}\right]^{(1+4\sigma-8\sigma^2)/3\Sigma},\label{dm4}\\
\frac{C_1 \sin[\sqrt{\Lambda}(R_1 -R_0)]}{\sqrt{\Lambda}}=\frac{\sinh(a_2 r_2)}{a_2[a_4 (\cosh(a_2 r_2)-1)+1]}
\left[\frac{Q(\rho_2)}{Q(\rho_1)}\right]^{-1/3}
\left[\frac{P(\rho_2)}{P(\rho_1)}\right]^{-2(1-2\sigma-2\sigma^2)/3\Sigma}.\label{dm5}
\end{align}
From the last two equations, one obtains
\begin{equation}\label{dm6}
\frac{C_1}{B_1}\tan[\sqrt{\Lambda}(R_1 -R_0)]=\frac{\sinh(a_2 r)}{a_2}\left[\frac{P(\rho_2)}{P(\rho_1)}\right]^{-(1-4\sigma^2)/\Sigma},
\end{equation}
which, together with
\eqref{m22} and \eqref{sig6}, implies
\begin{equation}\label{dm8}
\left(\frac{C_1}{B_1}\right)^2=\frac{1}{a_{2}^{2}}\left[\frac{P(\rho_2)}{P(\rho_1)}\right]^{-2(1-4\sigma^2)/\Sigma},
\end{equation}
In turn,  \eqref{dm3} and \eqref{dm8} give
\begin{equation}
A_1=\frac{\cosh(a_2 r_2)}{a_4 [\cosh(a_2 r_2)-1]+1}\left[\frac{Q(\rho_1)}{Q(\rho_2)}\right]^{1/3}\left(\frac{a_2 C_1}{B_1}\right)^{-(1-8\sigma+4\sigma^2)/3(1-4\sigma^2)},
\end{equation}
which tells us how the $A_1, B_1$ and $C_1$ parameters of the Einstein universe must be related as a consequence of the matching.

We summarize our analysis, as follows:
\begin{proposition}
Consider $\Lambda>0$ and $1/2\le a_4<1$. Then, there exists a cylindrically symmetric non-singular global $\Lambda$-vacuum spacetime resulting from the matching of LT, with  $0<\sigma<1/4$, to \eqref{EUniv} and \eqref{conf-flat}. Furthermore, given the interior three parameters, $a_2, a_4$ and $\Lambda$, the global space-time is uniquely determined.
\end{proposition}
 The resulting spacetime is globally cylindrically symmetric with an anisotropic fluid interior, an intermediate $\Lambda$-vacuum and a $\Lambda$-dust exterior given by the Einstein universe. This model can easily be changed into a globally toroidal spacetime by changing the variable $z\in\RR$ to a cyclic one $\psi\in[0,2\pi[$ in the three metrics considered here plus a rescaling in metric \eqref{conf-flat}. In this case, the constants $b$ and $B_1$ can be interpreted as angle defects (see also \cite{GP}).

\section{Circular geodesics}
\label{circular}
Circular geodesics in LT spacetimes for $\Lambda<0$ were studied in \cite{BDMS-2014} and for $\Lambda=0$ in \cite{Banerjee0}. In this section, we extend those studies to the case $\Lambda>0$.

\subsection{Tangential velocity and acceleration}
By restricting to study circular geodesics in the plane perpendicular to the $z$ axis we require
${\dot\rho}={\dot z}=0$.
The squared angular velocity of the particle around the $z$ axis is
$\omega^2=({\dot\phi}/{\dot t})^2=f^{\star}/l^{\star}$, where the star stands for $\rho$ differentiation, and the squared norm of its tangential velocity $W$ is then given by
$W^2=(l/f)\omega^2$ which, using \eqref{g2}, \eqref{g4} and \eqref{1}, gives
\begin{equation}
W^2=\frac{2\Sigma\sin^2R-6\sigma}{2\Sigma\sin^2R-3(1-2\sigma)}.\label{g11}
\end{equation}
This expression is non-singular for $0\le \sigma\le1/4$ and any $\rho$. Noting that, for $\sigma=0$ and $\sigma=1/2$, the right hand side of \eqref{g11} is negative, we deduce
\begin{lemma}
For $\sigma=0$ and $\sigma=1/2$ there are no circular orbits. For any other $0<\sigma<1/2$, there are open sets of values of $\rho$ and $\Lambda$ such that circular orbits always exist.
 \end{lemma}

When $\Lambda=0$, i.e. in the LC spacetime, (\ref{g11}) becomes
\begin{equation}
W_{LC}^{2}=\frac{2\sigma}{1-2\sigma}, ~\text{for}~ \sigma<1/2, \label{g11a}
\end{equation}
as expected (see \cite{BDMS-2014}). For a small enough $\Lambda$, one can expand \eqref{g11} in powers of $\Lambda$ to get
\begin{equation}
W^2= W_{LC}^2-\Lambda\frac{\Sigma}{2}\frac{1-4\sigma}{(1-2\sigma)^2}\rho^2+ O(\Lambda^2) , \label{g111} ~~\text{for}~~\sigma\ne \frac{1}{2},
\end{equation}
showing that for $0\le \sigma<1/4$, at linear order, both $\rho$ and $\Lambda$ decrease the corresponding tangential velocity for the LC circular geodesics. However, for $\sigma>1/4$, $\Lambda$ increases the tangential velocity $W_{LC}^2$, implying that the geodesics are spacelike.
By differentiating (\ref{g11}) with respect to $\rho$, we obtain
\begin{equation}
W^{2\star}=\frac{3(4\sigma-1)\Sigma\sqrt{3\Lambda}\sin(2R)}{[\Sigma \cos(2R)+2(1-2\sigma-2\sigma^2)]^2}.
\label{g11b}
\end{equation}

For positive $\Lambda$, this shows that $W^{2\star}$ is negative for $0\leq\sigma<1/4$, and the tangential velocity is a decreasing function of $\rho$, while $W^{2\star}$ is positive for
$\sigma>1/4$ and the tangential velocity is an increasing function of $\rho$. This is the opposite effect with respect to the $\Lambda<0$ case (see \cite {BDMS-2014}).

 This result for $0<\sigma<1/4$ can be understood as follows. Calculating the mass per unit length $m$ inside a cylindrical surface $S$ with centre at $\rho=0$ by using Israel's expression \cite{Israel}, which produces the Newtonian limit unlike other proposals, we obtain \cite{Brito}
\begin{equation}
m\stackrel{S}{=}m_{LC}-\frac{abc}{3}\sin^2R, \label{1}
\end{equation}
where
\begin{equation}
m_{LC}=abc\frac{\sigma}{\Sigma}, \label{2}
\end{equation}
is the mass per unit length produced by the LC spacetime. It is clear from (\ref{1}) that $\Lambda>0$ decreases the mass per unit length, as a consequence the circular geodesics have a smaller
%
tangential velocity producing a smaller
centrifugal force needed to withstand the gravitational attractive force of the source. While for $\Lambda<0$, since it increases the mass per unit length, the centrifugal force has to be bigger, thereby increasing its tangential velocity in order to withstand the gravitational force of the source as shown in \cite{BDMS-2014}.
For $\sigma=1/4$, equations (\ref{g11}) and (\ref{g11b}) become
\begin{equation}
W_{1/4}^{2}=1,\;\;\; W_{1/4}^{2\star}=0, \label{g12}
\end{equation}
which means that, independently of the $\rho$ distance and the value of $\Lambda$, the circular geodesics are null, like in the LC spacetime.

\begin{figure}[H]
\begin{minipage}{8cm}
\begin{tabular}{c}
\includegraphics[width=8cm]{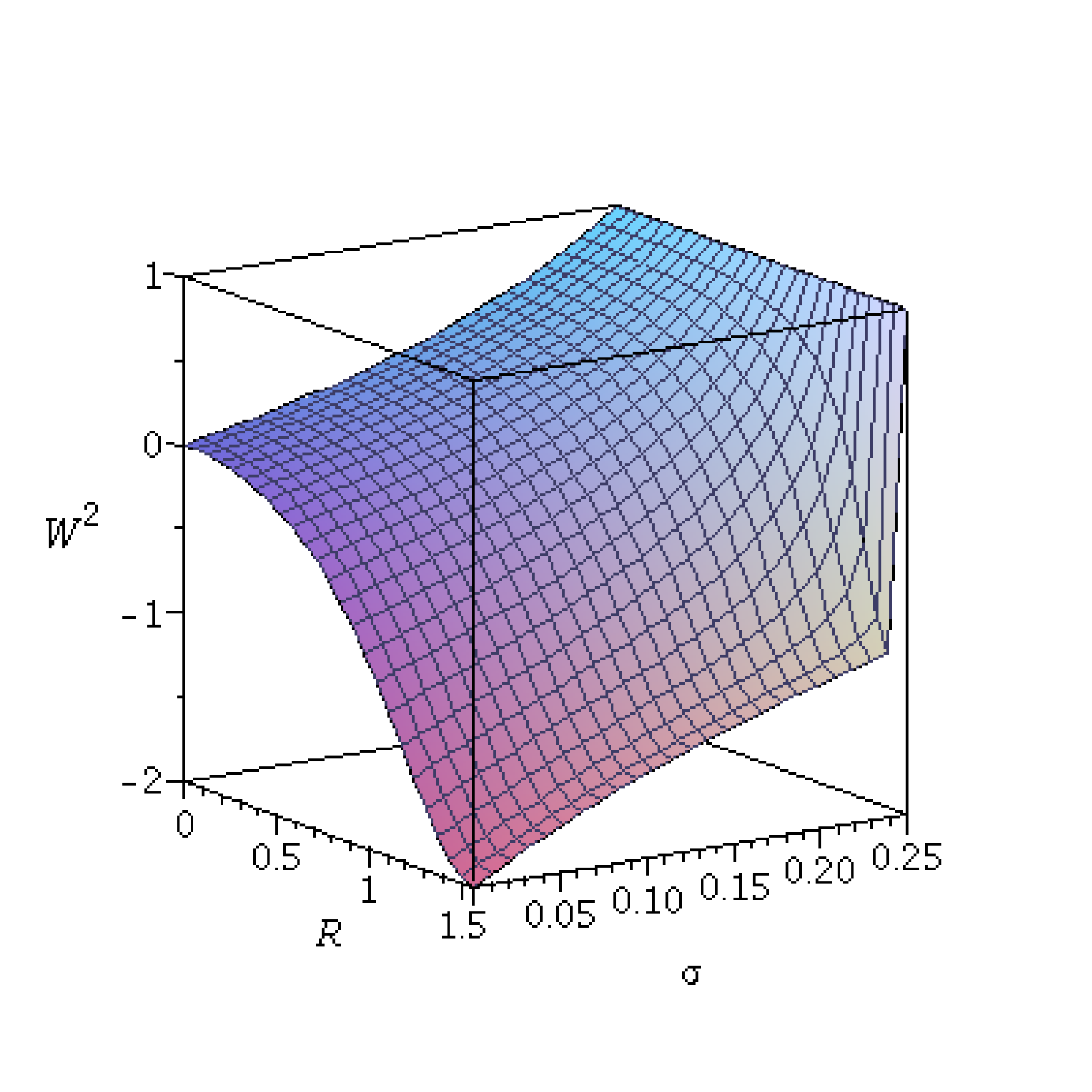} \\
\end{tabular}
\end{minipage}
\begin{minipage}{8cm}
\begin{tabular}{c}
\includegraphics[width=8cm]{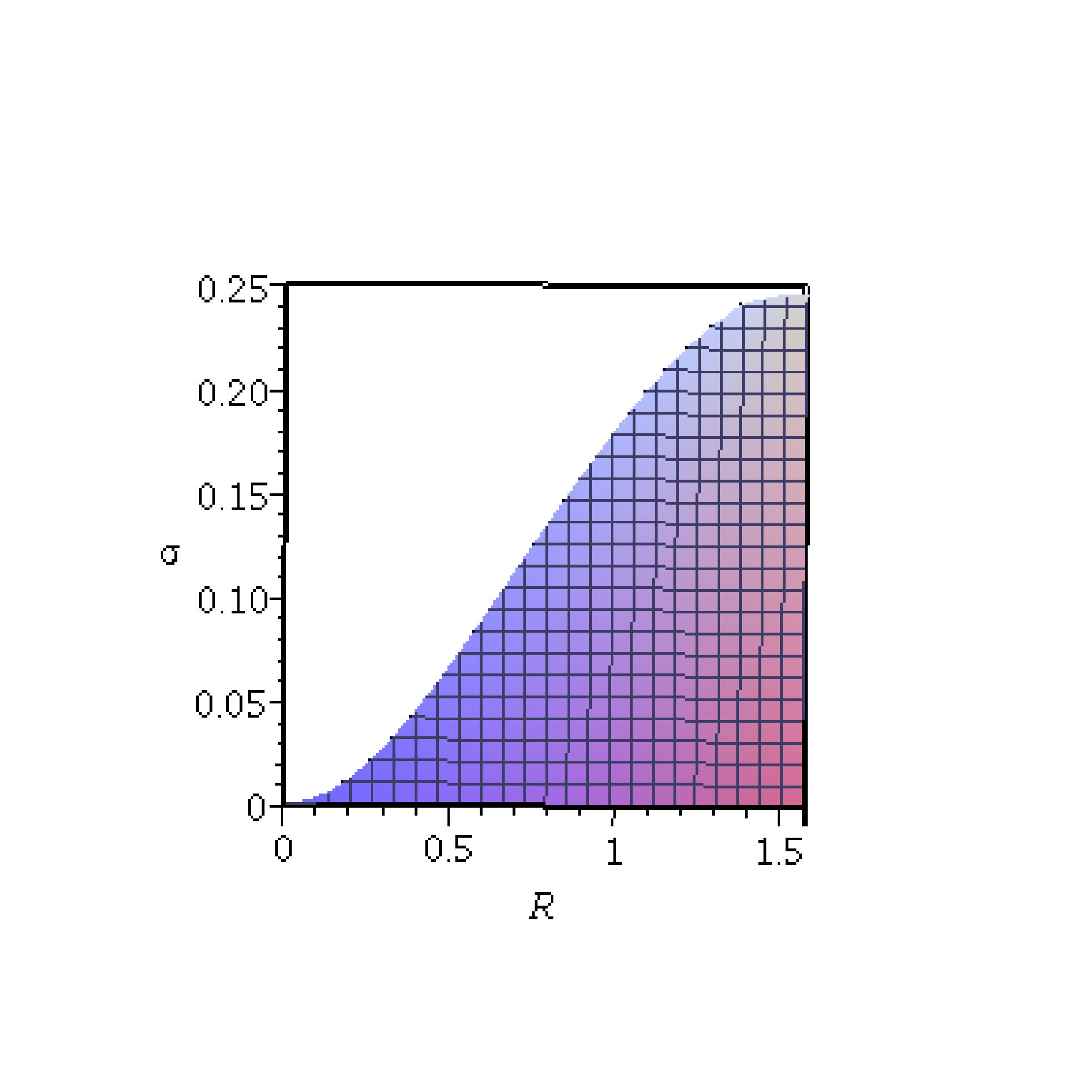}\\
\end{tabular}
\end{minipage}
\caption{ \textsl{Left panel}: Graphs of the squared tangential velocity $W^2$ for  $0\leq\sigma\leq 1/4$, and for values of $R$ within the interval $(0,\pi/2)$. \textsl{Right panel}: Graphs showing the regions where the right hand side of \eqref{g11} is positive (white)  and negative (colored) depending on $\sigma$ and $R$.}
\label{figW2}
\end{figure}

For $0 < \sigma < 1/4$ the geodesics are timelike but there is an upper limit for $\rho$ depending on the parameters $\sigma$ and $\Lambda$, so that circular geodesics can exist. This is represented in Figure \ref{figW2}.

%
\subsection{Proper radius}
\begin{figure}[H]
\begin{minipage}{8cm}
\begin{tabular}{c}
\includegraphics[width=8cm]{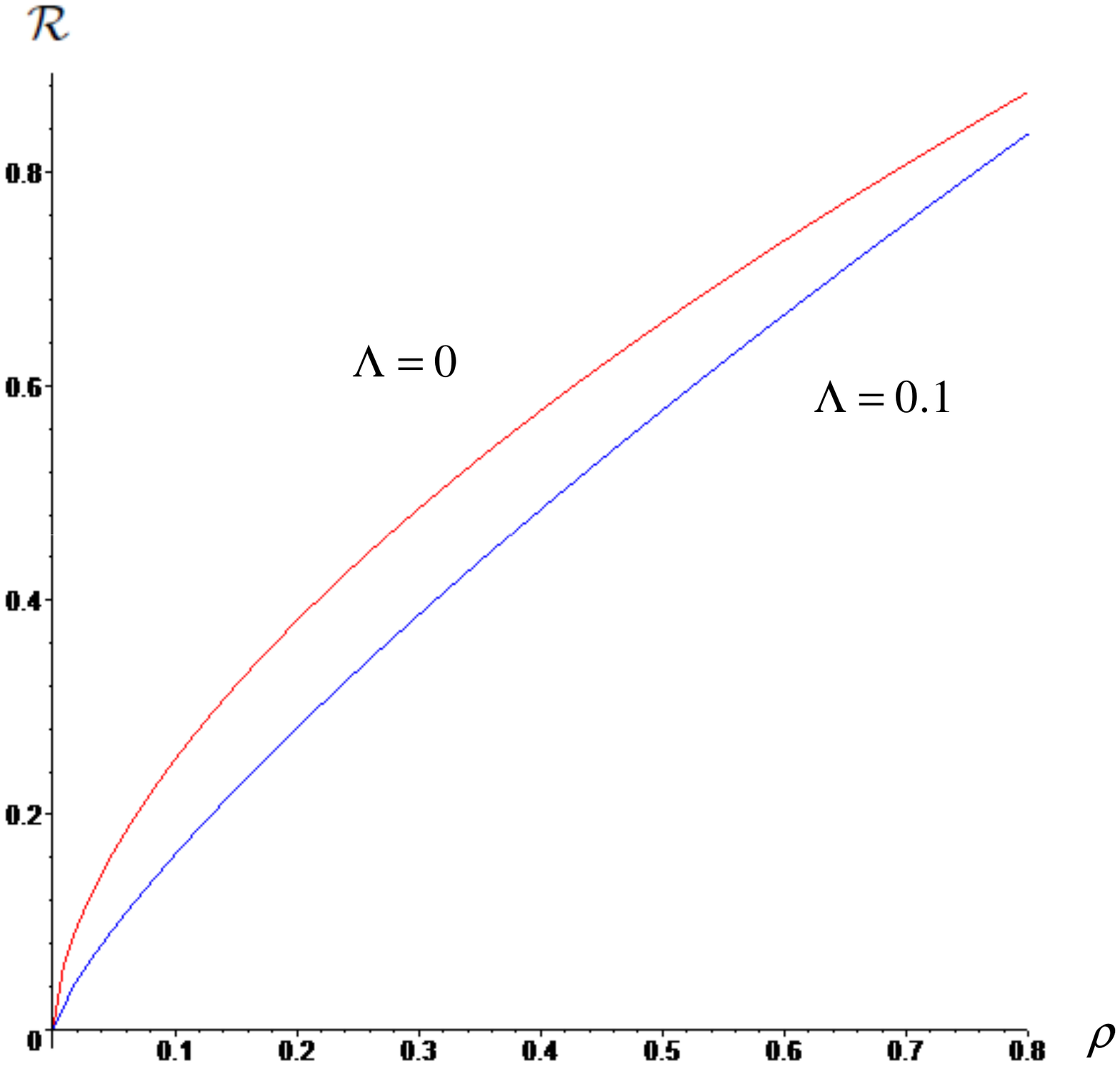} \\
\end{tabular}
\end{minipage}
\begin{minipage}{8cm}
\begin{tabular}{c}
\includegraphics[width=8cm]{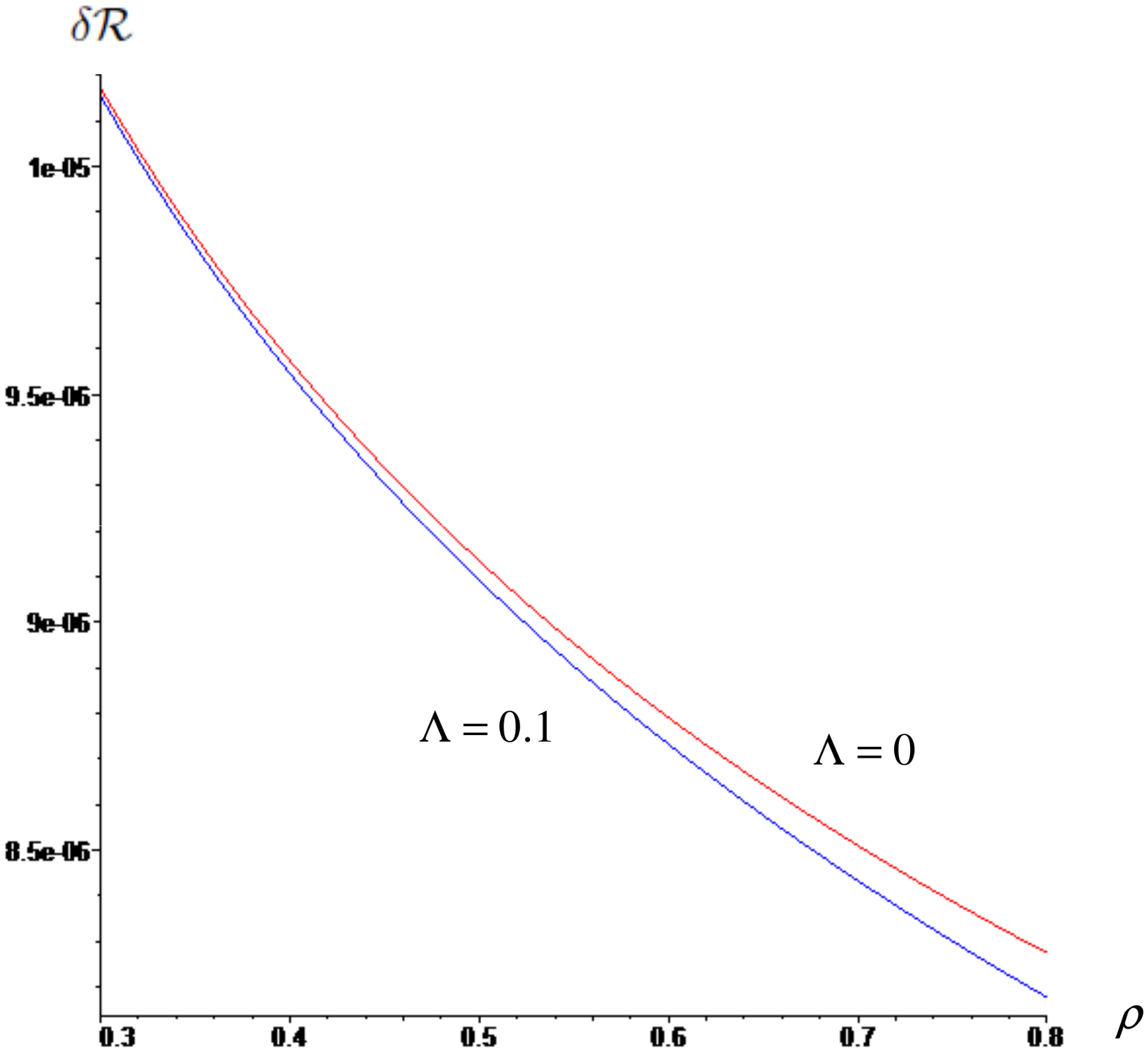}\\
\end{tabular}
\end{minipage}
\caption{ \textsl{Left panel}: Graphs of the proper radius $\mathcal{R}$ for $c =1$, $\sigma=1/5$, in the cases
$\Lambda=0$, where $\mathcal{R}=\sqrt{\rho^{2(1-2\sigma)}},$ and $\Lambda=0.1$. \textsl{Right panel}: Graphs of $\delta\mathcal{R}$ for $\Lambda=0$ and $\Lambda=0.1$, considering $\delta\rho =0.00001$. These figures illustrate $\Lambda$ stabilising $\mathcal{R}$ and $\delta\mathcal{R}$.}
\label{fig:pR}
\end{figure}

The proper radius $\mathcal{R}=\sqrt{g_{33}(\rho)}$, for any finite change $\delta\rho$ in the coordinate $\rho$, is given by
\begin{equation}
\delta\mathcal{R}=
\sqrt{g_{33}(\rho+\delta\rho)}-\sqrt{g_{33}(\rho)}. \label{g14}
\end{equation}
Directly from the LT metric, we obtain
\begin{eqnarray}
Q(\rho+\delta\rho)=Q(\rho)\left[1+\sqrt{3\Lambda}\frac{\cos(2R)}{\sin(2R)}\delta\rho\right], \label{g15}\\
P(\rho+\delta\rho)=P(\rho)\left[1+\sqrt{3\Lambda}\frac{\delta\rho}{\sin(2R)}\right], \label{g16}
\end{eqnarray}
together with
\begin{equation}
\sqrt{g_{33}(\rho+\delta\rho)}=\sqrt{g_{33}(\rho)}\left\{1+\sqrt{\frac{\Lambda}{3}}\left[\cos(2R)
+\frac{2(1-2\sigma-2\sigma^2)}{\Sigma}\right]\frac{\delta\rho}{\sin(2R)}\right\}. \label{g17}
\end{equation}
Substituting (\ref{g17}) into (\ref{g14}) we obtain
\begin{eqnarray}
\delta\mathcal{R}=
c\left(\frac{3 \Lambda}{4}\right)^{2\sigma^2/\Sigma}
\frac{(\tan R)^{4(1-\sigma)^2/3\Sigma}}{(\sin R)^{4/3}}
\left(-\frac{2}{3}\sin^2R+\frac{1-2\sigma}{\Sigma}\right)\delta\rho, \label{g18}
\end{eqnarray}
which shows that, with increasing $\rho$, the proper radius decreases. Examples of this dynamical behaviour are depicted in Figure \ref{fig:pR}.

For the case $\Lambda=0$, (\ref{g18}) reduces, as expected, to
\begin{equation}
\delta\mathcal{R}_{LC}
=\frac{c}{\Sigma}(1-2\sigma)\rho^{-4\sigma^2/\Sigma}\delta\rho. \label{g19}
\end{equation}

After an expansion up to first order in $\Lambda$, (\ref{g18}) becomes
\begin{eqnarray}
\delta{\mathcal R}\approx\delta{\mathcal R}_{LC}-\Lambda\frac{c\sigma^2}{\Sigma^2}[3(1-2\sigma)+8\sigma^2]
\rho^{2(1-2\sigma+2\sigma^2)/\Sigma}
\delta\rho, \label{30}
\end{eqnarray}
from where we see, explicitly, how $\Lambda$ decreases the proper distance along $\rho$.

For the LC spacetime there are no unbounded timelike orbits, because the infinite line source is so strong, even for small $\sigma$. Since the proper radius is not increased asymptotically by increasing the coordinate radius $\rho\rightarrow\infty$, we have $\delta{\mathcal R}_{LC}\rightarrow 0$, as observed by Griffiths and Podolsky \cite{Griffiths}. The introduction of $\Lambda>0$, which diminishes the energy per unit length content of the source, intuitively should increase the extension of the proper radius, following the LC result. However, the result is just the opposite, as both $\Lambda$ and $\rho$ diminish the extent of the proper radius.
This shows how dramatic is the modification of the spacetime geometry by introducing $\Lambda$.

Comparing this result to the case $\Lambda<0$ studied in \cite{BDMS-2014}, we have that it is exactly the inverse result since the proper radius is increased by $\Lambda<0$ which increases the energy per unit length content of the source.
\section{Geodesic motion along $z$}
In this case, we have from (\ref{gg7})
\begin{eqnarray}
{\dot z}=\frac{P_z}{b^2} \frac{P^{2\zeta/3}}{Q^{2/3}},\label{g200}\\
{\ddot z}=\frac{2}{3}P_z
\left[\zeta-\cos(2R)\right]\frac{P^{2\zeta/3}}{Q^{5/3}}{\dot\rho}, \label{g20}
\end{eqnarray}
where
$$
\zeta= \frac{1+4\sigma-8\sigma^2}{\Sigma}.
$$
Since $\dot z>0$, and taking into account similar results for $\Lambda\le 0$, see \cite{BDMS-2014}, we get the following result:

\begin{lemma}
For any $\Lambda$, all geodesics with $P_z\ne 0$ are unbounded along $z$, unless they reach the spacetime singularity $\rho=0$ for a finite value of the affine parameter $\lambda$. In particular, for $\Lambda>0$, such geodesics never reach the singularity $\rho=\pi/\sqrt{3\Lambda}$.
\end{lemma}

If $\Lambda=0$, then \eqref{g200}-(\ref{g20}) reduce to
\begin{eqnarray}
{\dot z}=\frac{P_z}{b^2} \rho^{4\sigma(1-2\sigma)/\Sigma}, \label{g210}\\
{\ddot z}=P_z\frac{4\sigma(1-2\sigma)}{\Sigma}
\frac{\dot\rho}{\rho^{(1-6\sigma+12\sigma^2)/\Sigma}}. \label{g21}
\end{eqnarray}
So, for $\Lambda\ge 0$ we can see that, for $0<\sigma<1/2$ and $P_z>0$, the particle along the $z$ direction tends always to accelerate (decelerate) for $\dot\rho>0$ ($\dot\rho<0$), i.e. for increasing (decreasing) radial distances from the axis. In fact, $\dot z\to \infty$, as $\rho\to \pi/\sqrt{3\Lambda}$.
From \eqref{gg7}, we easily see that for $\Lambda>0$ and $0<\sigma<1/2$, the velocity $\dot z \to 0$ as $\rho\to 0$.

For $\sigma=0$ or $\sigma=1/2$, and $\Lambda>0$, we obtain
\begin{eqnarray}
{\dot z}=\frac{P_z}{b^2 (\cos R)^{4/3}}, \label{g20a0}\\
{\ddot z}=\frac{2}{3}P_z\left[1-\cos(2R)\right]\frac{P^{2/3}}{Q^{5/3}}{\dot\rho}, \label{g20a}
\end{eqnarray}
which shows that for $\rho\to 0$ we get $\dot z\to P_z/b^2$ whereas for $\rho\to \pi/\sqrt{3\Lambda}$, we again get $\dot z\to+\infty$. If $\Lambda=0$, the previous expressions reduce to $\ddot z=0$ and $\dot z=P_z/ b^2 $.

Hence, physically, we see that the inclusion of $\Lambda>0$ does not alter the main features of the purely relativistic properties of the acceleration along the $z$ axis due to the displacement of a test particle along the radial distance $\rho$, as seen in (\ref{g21}). By contrast, the introduction of $\Lambda<0$ changes radically the behaviour of (\ref{g21}), as seen in \cite{BDMS-2014}.

If $\dot\rho=0$, then $\ddot z=0$ and the geodesics describe circular helices, except if $L_z=0$ which corresponds to straight lines along the $z$ direction, or if $P_z=0$ which corresponds to circles on planes perpendicular to the $z$ axis (see the previous section).

\begin{figure}[H]
\begin{minipage}{8cm}
\begin{tabular}{c}
\includegraphics[width=8cm]{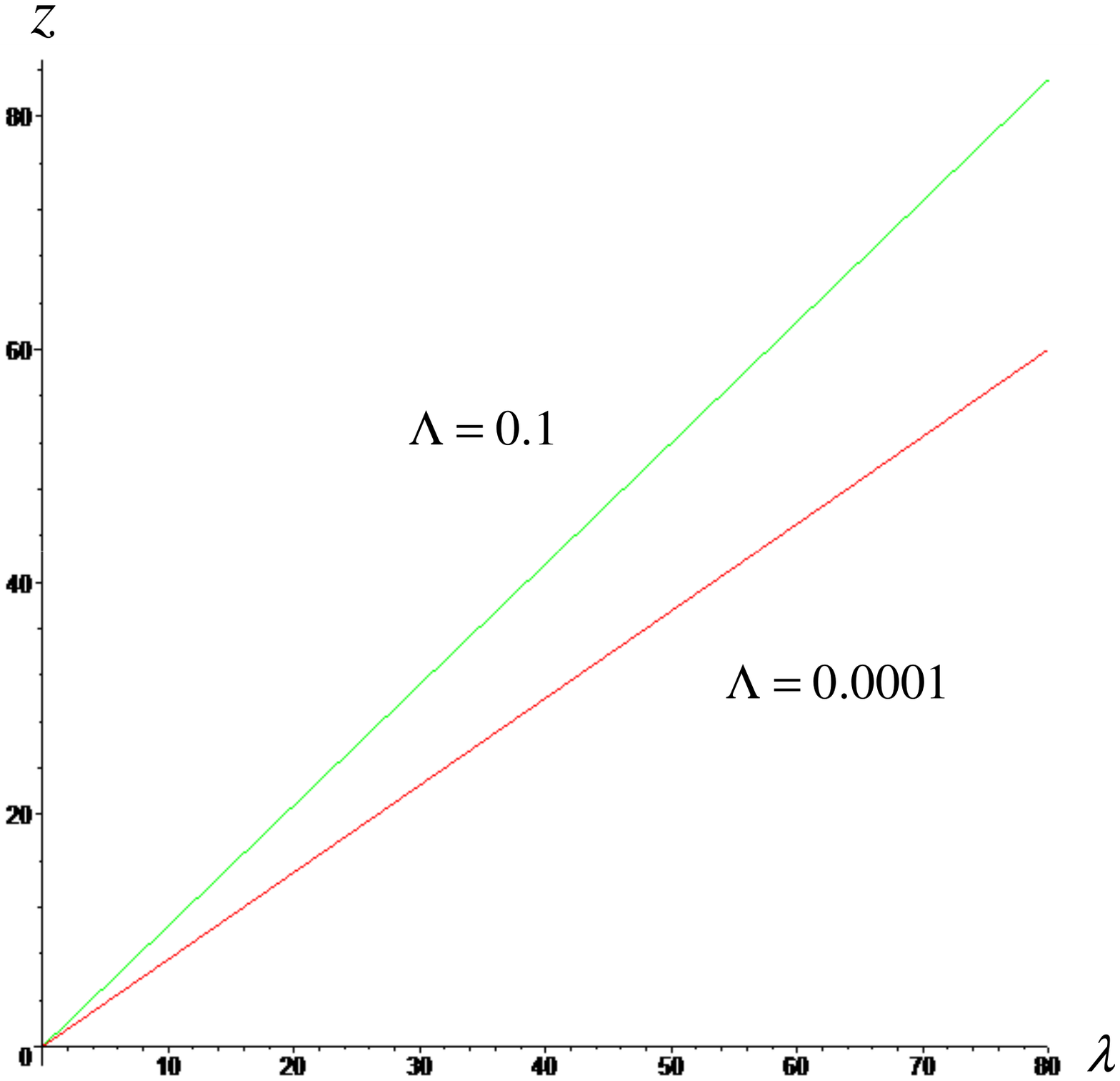} \\
\end{tabular}
\end{minipage}
\begin{minipage}{8cm}
\begin{tabular}{c}
\includegraphics[width=8cm]{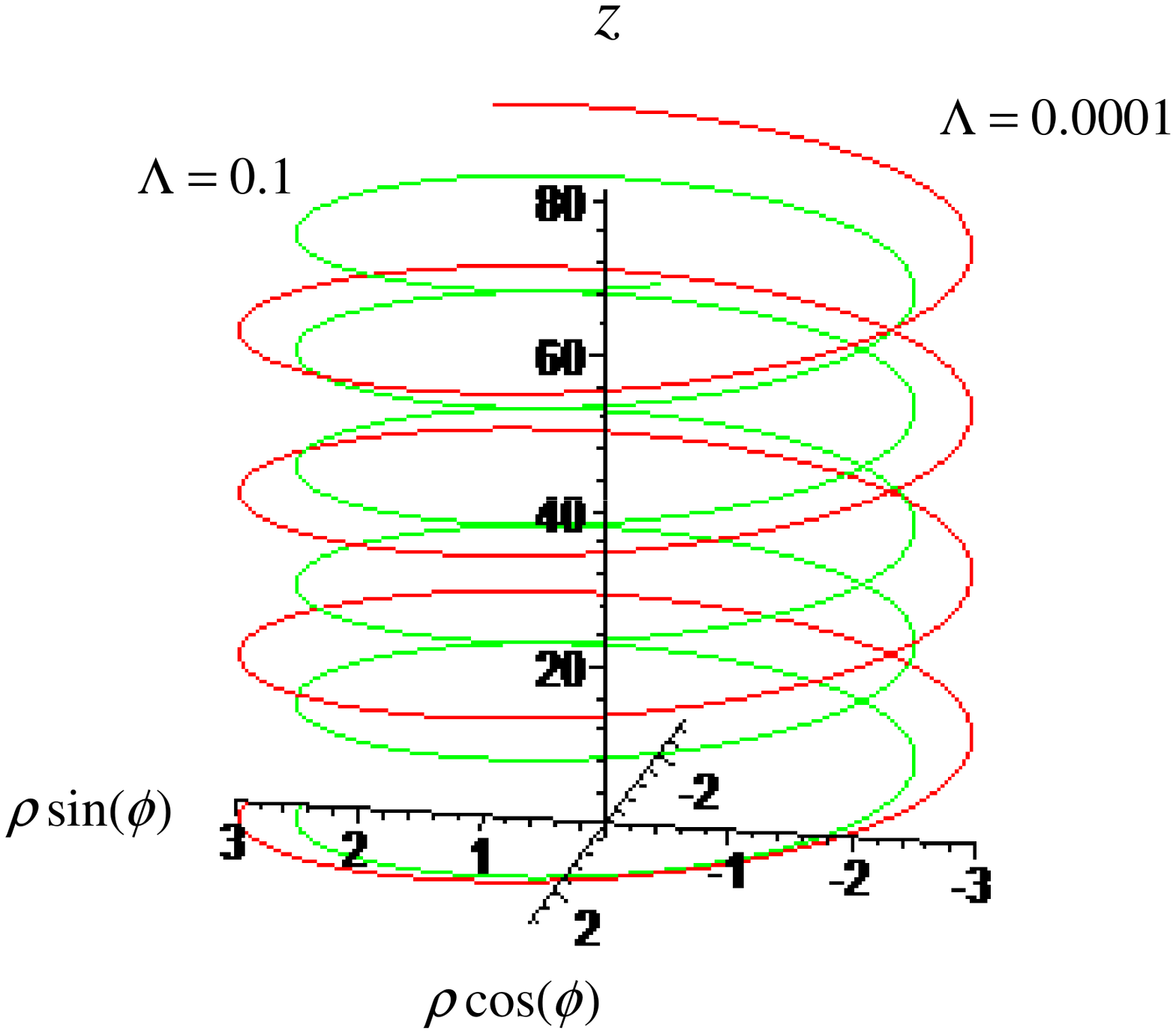}\\
\end{tabular}
\end{minipage}
\caption{ Graphs of the numerical integration of the geodesics' equations along $z(\lambda)$ in the case $\dot{\rho}=0$, for $E=2$, $\epsilon=0$, $b=c=1$, $P_z =L_z=0.5,$ $\sigma=0.4$, for $\Lambda=0.0001$ (for which $\rho=2.9014$) and for $\Lambda=0.1$ (for which $\rho=2.4521$).}
\label{fig:geodh}
\end{figure}

\begin{figure}[H]
\begin{minipage}{8cm}
\begin{tabular}{c}
\includegraphics[width=8cm]{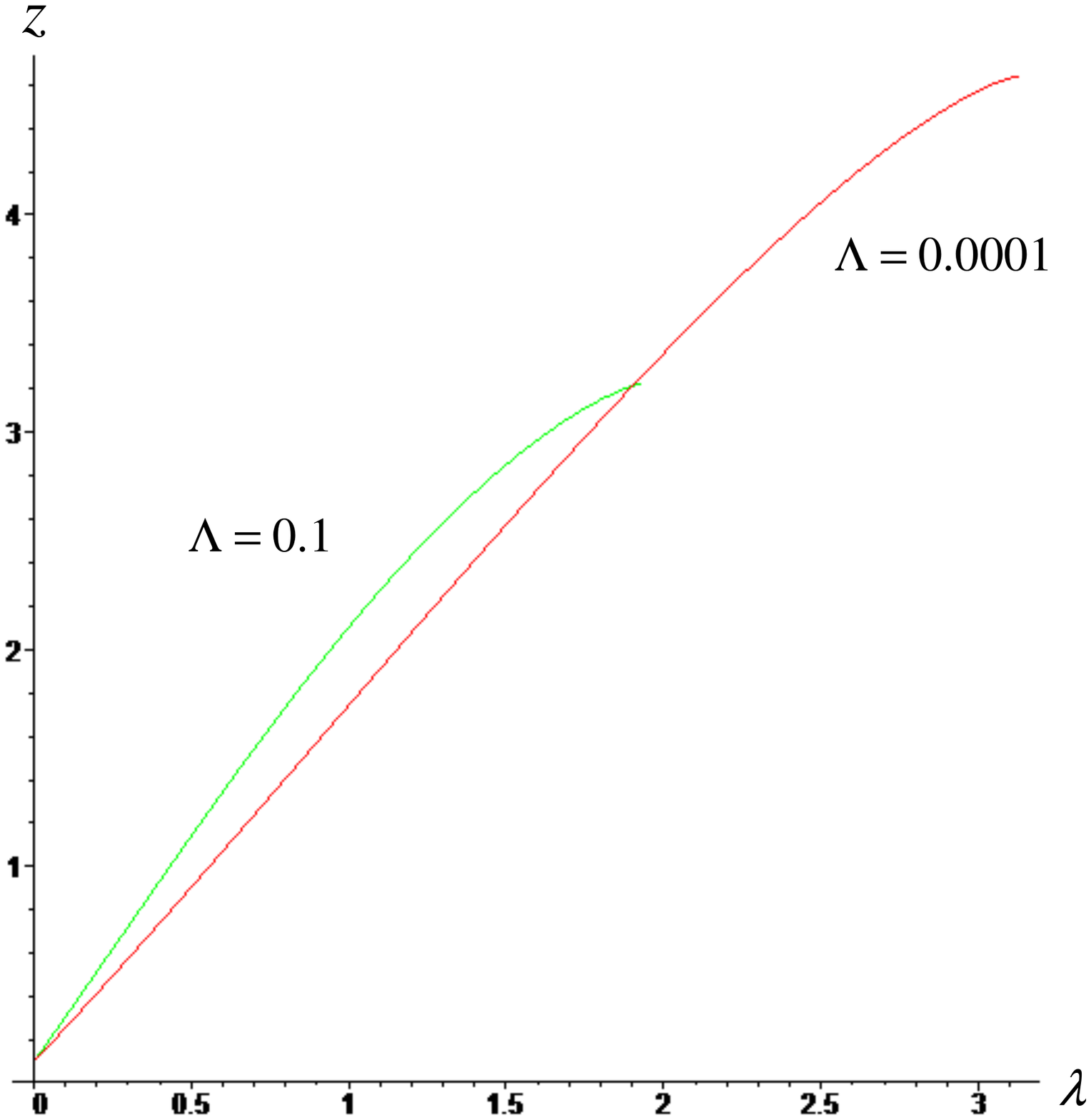} \\
\end{tabular}
\end{minipage}
\begin{minipage}{8cm}
\begin{tabular}{c}
\includegraphics[width=8cm]{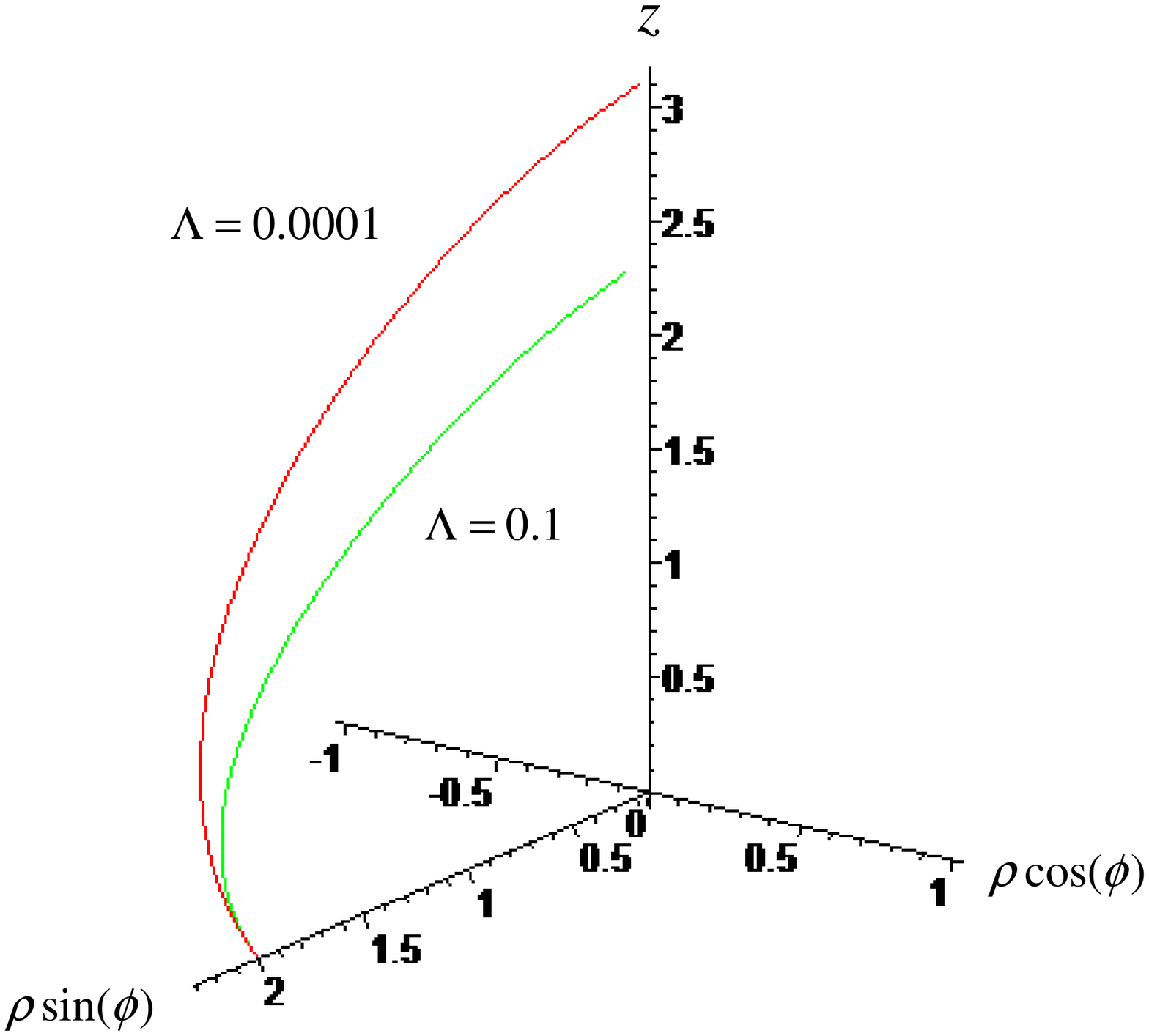}\\
\end{tabular}
\end{minipage}
\caption{ Graphs of the numerical integration of the geodesics' equations along $z(\lambda)$, for $E=2$, $\epsilon=L_z=0$, $b=P_z=1$, $\sigma=0.2$, for $\Lambda=0.0001$ and for $\Lambda=0.1$.}
\label{fig:geod11}
\end{figure}

Numerical examples of the geodesics dynamics can be seen in Figures \ref{fig:geodh}-\ref{fig:geod5} and Figures \ref{fig:geod6}-\ref{fig:geod9}.

\section{Geodesics along $\rho$}
From (\ref{gg6}), as in the $\Lambda<0$ case, we can derive
\begin{equation}
{\dot\rho}^2=Q^{-2/3}P^{2(1-8\sigma+4\sigma^2)/3\Sigma}[E^2-V(\rho)], \label{23}
\end{equation}
where
\begin{equation}
V(\rho)=\epsilon Q^{2/3}P^{-2(1-8\sigma+4\sigma^2)/3\Sigma}+\left(\frac{P_z}{b}\right)^2 P^{8\sigma(1-\sigma)/\Sigma}+\left(\frac{L_z}{c}\right)^2
P^{-2(1-4\sigma)/\Sigma} \label{24}
\end{equation}
is positive for $\epsilon = 0, 1$, i.e. for null or timelike geodesics. These two quantities will be crucial to analyse the dynamics of the geodesics in each case ahead.
\begin{lemma} \label{lemma-estimates}From (\ref{23}) and (\ref{24}), we get the following estimates:
\begin{itemize}
\item As
$\rho\rightarrow 0$, we have that:
\begin{itemize}
\item [$\bullet$]
$V(\rho)\rightarrow\infty$, if $\sigma<1/4$ and $L_z\ne 0$;\\
$V(\rho)\rightarrow 0$, if $L_z=0$ and $\sigma\ne 0$ or if $L_z\ne 0$ and $\sigma>1/4$;\\
$V(\rho)\rightarrow (L_z/c)^2$, for $\sigma=1/4$;\\
$V(\rho)\rightarrow \epsilon+ (P_z/b)^2$, if $L_z=0$ and $\sigma=0$.\\
\item [$\bullet$] $\dot\rho^2\to +\infty$, for $\sigma>0$;\\
$\dot\rho^2\to E^2-[\epsilon+ (P_z/b)^2]$, for $\sigma=0$ and $L_z=0$.
\end{itemize}
\item As
$\rho\rightarrow \pi/\sqrt{3\Lambda}$, we have that:
\begin{itemize}
\item [$\bullet$]  $V(\rho)\rightarrow\infty$, if $\sigma>1/4$ and for any $P_z$ or if $0<\sigma\le 1/4$ and $P_z\ne 0$;\\
$V(\rho)\rightarrow 0$, if $\sigma<1/4$ and $P_z=0$;\\
$V(\rho)\rightarrow \epsilon(2/\sqrt{3\Lambda})^{4/3}+ (L_z/c)^2$, for $\sigma=1/4$ and $P_z=0$;\\ $V(\rho)\rightarrow (P_z/b)^2$, if $\sigma=0$.\\
\item [$\bullet$] $\dot\rho^2\to +\infty$, for $0\le\sigma<1/4$;\\
$\dot\rho^2\to 0$, for $\sigma>1/4$;\\
$\dot\rho^2\to (2/\sqrt{3\Lambda})^{-4/3}[E^2-\epsilon(2/\sqrt{3\Lambda})^{4/3}-(L_z/c)^2]$, if $\sigma=1/4$ and $P_z=0$.
\end{itemize}
\end{itemize}
\end{lemma}
These estimates will be used to investigate the admisable orbits in each case ahead, taking into account equation \eqref{23}. In order to do that we will also need the first and second order derivatives of $V$.
From (\ref{24}), we obtain
\begin{eqnarray}
V^{\star}(\rho)=\frac{2\sqrt{3\Lambda}}{\Sigma \sin(2R)}\left\{\frac{\epsilon}{3}\;Q^{2/3}P^{-2(1-8\sigma+4\sigma^2)/3\Sigma}
\left[\Sigma \cos(2R)-1+8\sigma-4\sigma^2\right]\right. \nonumber\\
\left.+4\sigma(1-\sigma)\left(\frac{P_z}{b}\right)^2P^{8\sigma(1-\sigma)/\Sigma}-
(1-4\sigma)\left(\frac{L_z}{c}\right)^2P^{-2(1-4\sigma)/\Sigma}\right\}, \label{25}
\end{eqnarray}
and
\begin{eqnarray}
V^{\star\star}(\rho)=-\frac{\sqrt{3\Lambda}V^{\star}(\rho)}{\tan(2R)}+\left[\frac{2\sqrt{3\Lambda}}{\Sigma \sin(2R)}\right]^2
\left\{- \frac{\epsilon}{6}\Sigma^2Q^{2/3}P^{-2(1-8\sigma+4\sigma^2)/3\Sigma} \sin^2(2R)\right.\nonumber\\
\left.+\frac{\epsilon}{9}Q^{2/3}P^{-2(1-8\sigma+4\sigma^2)/3\Sigma}\left[\Sigma \cos(2R)-1+8\sigma-4\sigma^2\right]^2\right. \nonumber\\
\left.+16\sigma^2(1-\sigma)^2\left(\frac{P_z}{b}\right)^2P^{8\sigma(1-\sigma)/\Sigma}+(1-4\sigma)^2\left(\frac{L_z}{c}\right)^2
P^{-2(1-4\sigma)/\Sigma}\right\}. \label{26}
\end{eqnarray}
In order to analyse the signs of  $V^\star$ and $V^{\star\star}$, it is useful to note that for $0\le\sigma\le 1/2$ and $0\le R\le \pi/2$ we have
\begin{equation}
\label{ineq1}
\frac{3}{4}\le\Sigma\le 1~~{\text {and}}~~-2\le  \Sigma\cos{(2R)}-1+8\sigma-4\sigma^2\le 3,
\end{equation}
and, for $\sigma>1/4$ and $0\le R\le \pi/2$, we have
\begin{equation}
\label{ineq2}
 \Sigma\cos{(2R)}-1+8\sigma-4\sigma^2>0.
\end{equation}
When the equation $V^{\star}(\rho)=0$ has a solution, say $\rho=\rho_e$, satisfying
\begin{eqnarray}
(1-4\sigma)\left(\frac{L_z}{c}\right)^2P^{-2(1-4\sigma)/\Sigma}\stackrel{\rho_e}{=}
\frac{\epsilon}{3}\;Q^{2/3}P^{-2(1-8\sigma+4\sigma^2)/3\Sigma}
\left[\Sigma \cos(2R)-1+8\sigma-4\sigma^2\right] \nonumber\\
+4\sigma(1-\sigma)\left(\frac{P_z}{b}\right)^2 P^{8\sigma(1-\sigma)/\Sigma}, \label{27}
\end{eqnarray}
we obtain, after substituting the $L_z$ term into (\ref{26})
\begin{eqnarray}
V^{\star\star}(\rho_e)=\left[\frac{2\sqrt{3\Lambda}}{\Sigma \sin(2R)}\right]^2\left\langle\frac{\epsilon}{3}Q^{2/3}
P^{-2(1-8\sigma+4\sigma^2)/3\Sigma}\left\{-\frac{\Sigma^2}{2}\sin^2(2R)\right.\right.\nonumber\\
\left.\left.+\frac{1}{3}\left[\Sigma \cos(2R)-1+8\sigma
-4\sigma^2\right]^2
+(1-4\sigma)\left[\Sigma \cos(2R)-1+8\sigma-4\sigma^2\right]\right\}\right.\nonumber\\
\left.+4\sigma(1-4\sigma^2)(1-\sigma)\left(\frac{P_z}{b}\right)^2P^{8\sigma(1-\sigma)/\Sigma}\right\rangle. \label{28}
\end{eqnarray}
The equation $V(\rho)=E^2$ allows to find the minimum or maximum distances, $\rho=\rho_{min}$ or $\rho=\rho_{max}$, when they exist, reached by a particle from the axis. In the case $\Lambda=0$, that equation reduces to
\begin{equation}
\epsilon \rho_{LCm}^{4\sigma/\Sigma}+ \left(\frac{P_{z}}{b}\right)^{2} \rho_{LCm}^{8\sigma(1-\sigma)/\Sigma}+\left(\frac{L_z}{c}\right)^2 \rho_{LCm}^{-2(1-4\sigma)/\Sigma}=E^2\label{38A}
\end{equation}
and, up to first order in $\Lambda$, gives
\begin{eqnarray}
\label{rhomrholcm}
\rho_{m}\approx\rho_{LCm} +\;\frac{\Lambda}{4}\; \left[\left(\frac{L_z}{c}\right)^2 (1-4\sigma) \rho_{LCm}^{4\sigma(1+2\sigma)/\Sigma}\right. \nonumber\\
\left.+ \epsilon (1-2\sigma)^2 \rho_{LCm}^{2(1+4\sigma^2 )/ \Sigma}- 4\left(\frac{P_{z}}{b}\right)^2 \sigma(1-\sigma)\rho_{LCm}^{2(1+2\sigma)/\Sigma} \right] \nonumber\\
\times
\left[-\left(\frac{L_z}{c}\right)^2 (1-4\sigma) \rho_{LCm}^{(-3+10\sigma-4\sigma^2)/\Sigma} +2\epsilon \sigma \rho_{LCm}^{(-1+6\sigma-4\sigma^2)/\Sigma} \right.\nonumber\\
\left.+4\left(\frac{P_{z}}{b}\right)^{2}\sigma(1-\sigma)\rho_{LCm}^{(-1+10\sigma-12\sigma^2)/\Sigma} \right]^{-1},\label{38B}
\end{eqnarray}
where $\rho_{LCm}$ and $\rho_{m}$ denote the extreme (minimum or maximum) values of $\rho$ in the LC and LT spacetimes, respectively. From this relation, we
will be able to conclude, in some cases, that $\Lambda$ decreases the extreme distances of the geodesics to the axis. Although (\ref{38B}) is approximate, we will obtain an exact relation between $\rho_m$ and $\rho_{LCm}$ in the particular cases $\epsilon=P_z=0$ and $\epsilon=L_z=0$.

If $\Lambda=0$, we recall that (\ref{23}) becomes
\begin{equation}
{\dot\rho}^2_{LC}=\left[E^2- \epsilon \rho^{4\sigma/\Sigma}-\left(\frac{P_{z}}{b}\right)^{2} \rho^{8\sigma(1-\sigma)/\Sigma}-\left(\frac{L_z}{c}\right)^2 \rho^{-2(1-4\sigma)/\Sigma} \right]\rho^{-4\sigma/\Sigma},
\label{38C}
\end{equation}
and expanding in $\Lambda$, we obtain from (\ref{23}) and (\ref{38C})
\begin{equation}
{\dot\rho}^2\approx{\dot\rho}^2_{LC}+\frac{\Lambda}{2\Sigma}\left[-4\sigma^2\left(\frac{L_z}{c}\right)^2\rho^{8\sigma^2/\Sigma}
+(1-2\sigma)^2E^2\rho^{2(1-2\sigma)^2/\Sigma}-\left(\frac{P_{z}}{b}\right)^{2} \rho^{2/\Sigma}\right].
\end{equation}

For the acceleration, we get
\begin{eqnarray}
{\ddot\rho}=-\frac{P^{2(1-8\sigma+4\sigma^2)/3\Sigma}}{3\Sigma Q^{5/3}}\left[\Sigma \cos(2R)-1+8\sigma-4\sigma^2\right]
\nonumber\\
\times\left\{E^2-\left(\frac{P_z}{b}\right)^2 \left[\frac{\Sigma \cos(2R)-1-4\sigma+8\sigma^2}{\Sigma \cos(2R)-1+8\sigma-4\sigma^2}\right]
P^{8\sigma(1-\sigma)/\Sigma}\right. \nonumber\\
\left.-\left(\frac{L_z}{c}\right)^2\left[\frac{\Sigma \cos(2R)+2(1-2\sigma-2\sigma^2)}
{\Sigma \cos(2R)-1+8\sigma-4\sigma^2}\right]P^{-2(1-4\sigma)/\Sigma}\right\}. \label{29}
\end{eqnarray}
The limits of $V^\star, V^{\star\star}$ and  $\ddot\rho$, as $\rho\to 0$ and as $\rho\to \pi/(\sqrt{3\Lambda})$, can be written down but depend non-trivially on $\sigma$, so we omit them here and refer to them later when necessary.

The cases where $E^2< V$ give $\dot\rho^2<0$, which is physically not acceptable and we omit them here. In turn, the cases where $E^2=V$ give $\dot\rho^2=0$ and correspond to circular helices (along cylindrical surfaces) or to planar circles (if $P_z=0$), which have been studied in Section \ref{circular}.

As in \cite{BDMS-2014}, we now split our analysis into planar and non-planar as well as null and non-null
geodesics, and use the above formulae to study their dynamics, in each case.
\subsection{Planar geodesics ($P_z=0$)}
Planar geodesics for $\Lambda<0$ were studied in \cite{Banerjee,BDMS-2014}, where bounded orbits were found in some cases, and their stability was carefully analysed. In this section, besides extending this analysis to $\Lambda>0$, we do a comparative study with the cases $\Lambda=0$ and $\Lambda<0$. Throughout, we consider $E^2>V$.
\subsubsection{Case $\epsilon=0$}
In this case,  (\ref{24}) becomes
\begin{equation}
V(\rho)=\left(\frac{L_z}{c}\right)^2P^{-2(1-4\sigma)/\Sigma}, \label{30}\\
\end{equation}
and, from (\ref{25}), one can conclude that $V^{\star}(\rho)<0$ if $\sigma<1/4$, $V^{\star}(\rho)=0$ if $\sigma=1/4$ and $V^{\star}(\rho)>0$ if $\sigma>1/4$. Therefore, we study the following three cases:
\begin{enumerate}
\item{$\sigma<1/4$}

Null particles approaching $z$ have decreasing negative acceleration, $\ddot\rho<0$, and increasing speed $\dot\rho$. The particles reach their minimum distance from the axis for
    \begin{equation}
    P_{min}=\left(\frac{L_z}{cE}\right)^{\Sigma/(1-4\sigma)}, \label{46a}
    \end{equation}
    from which we can extract $\rho_{min}$. One should bare in mind that the minimum (respectively maximum) distance to $\rho=0$ corresponds to the maximum (respectively minimum) distance to $\rho=\pi/\sqrt{3\Lambda}$. For $\Lambda=0$, we have from (\ref{46a})
    \begin{equation}
    \rho_{LCmin}=\left(\frac{L_z}{cE}\right)^{\Sigma/(1-4\sigma)}, \label{46b}
    \end{equation}
    which is the minimum distance from the $z$ axis attained by an incoming null particle in the LC spacetime.
    From (\ref{46a}) and (\ref{46b}), we have
    \begin{equation}
    \frac{\sqrt{3\Lambda}}{2}\,\rho_{LCmin}=\tan\left(\frac{\sqrt{3\Lambda}}{2}\,\rho_{min}\right), \label{46bb}
    \end{equation}
implying  $\rho_{min}<\rho_{LCmin}$ and that $\Lambda$ decreases the minimum distance $\rho_{min}$ of the null particle to the $\rho=0$ axis. This is the opposite behaviour with respect to the $\Lambda<0$ case. A linear version of this result, can immediately be derived from  \eqref{38B} and non-linear numerical examples are plotted in Figure \ref{fig:geod3}. From its minimum distance from the axis, the particle reaches the second singularity with increasing speed and increasing acceleration.

\begin{figure}[H]
\begin{minipage}{8cm}
\begin{tabular}{c}
\includegraphics[width=8cm]{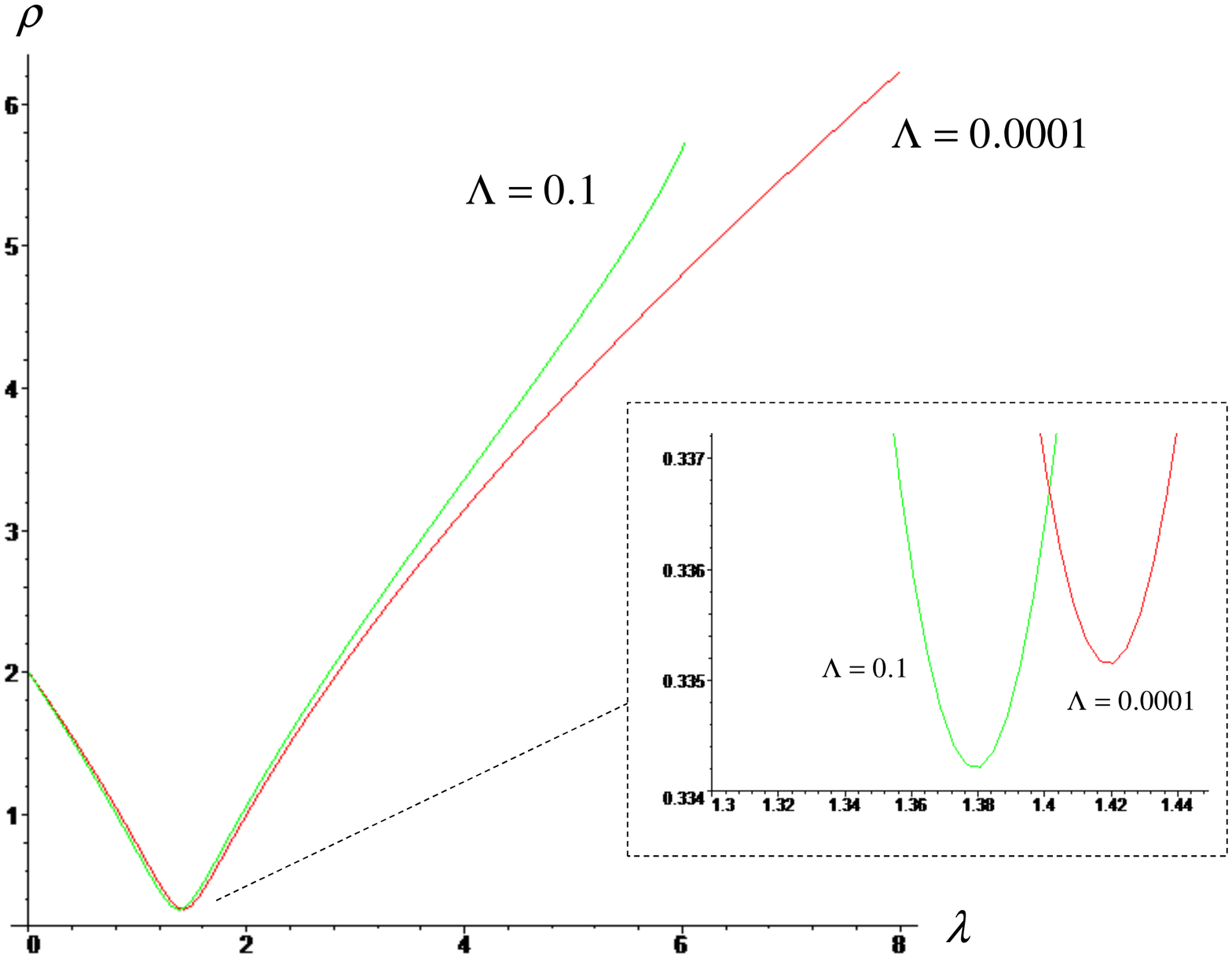} \\
\end{tabular}
\end{minipage}
\begin{minipage}{8cm}
\begin{tabular}{c}
\includegraphics[width=8cm]{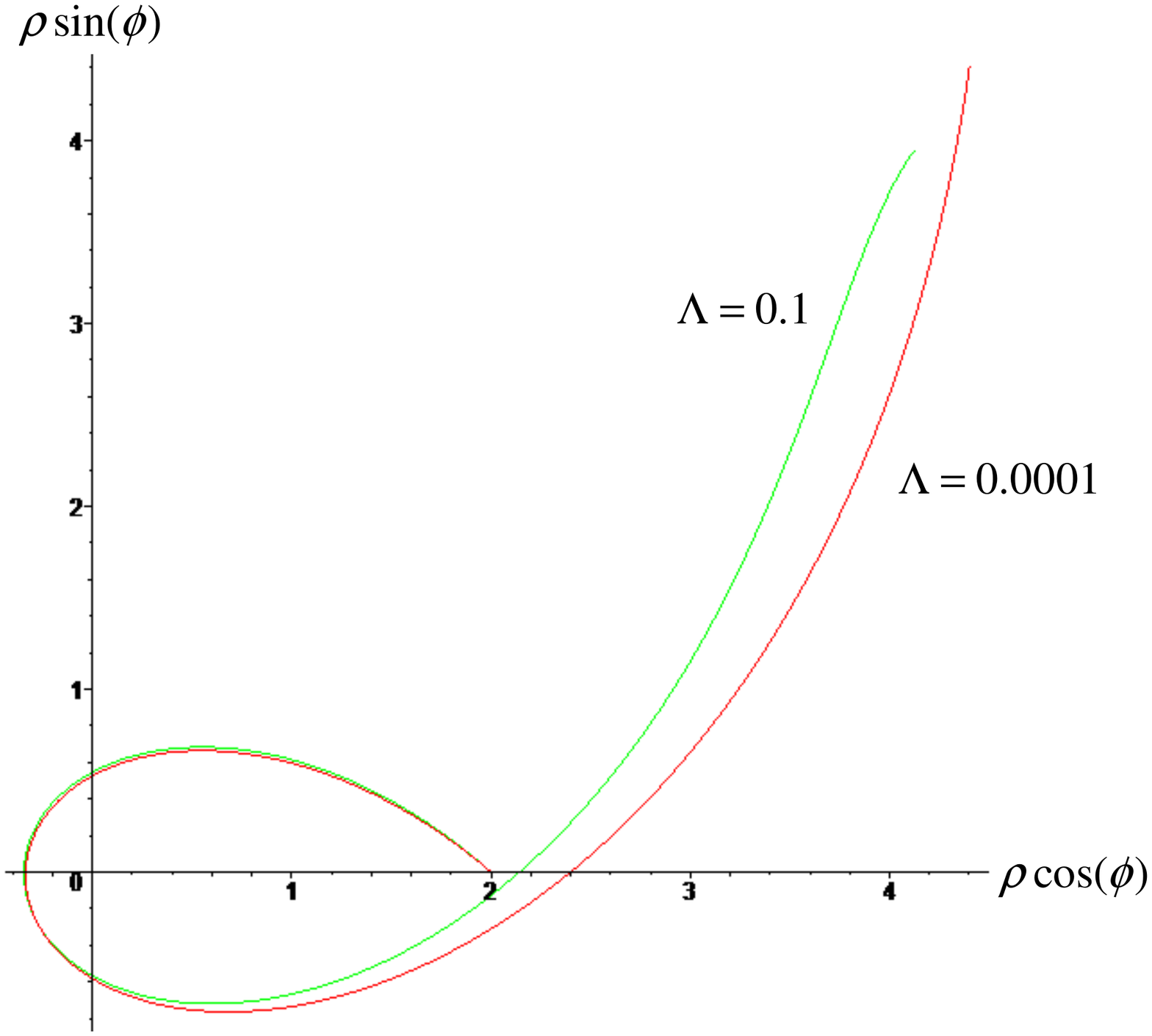}\\
\end{tabular}
\end{minipage}
\caption{Graphs of the numerical integration of the geodesics' equations along $\rho(\lambda)$, for $E=2$, $b=c=1$, $L_z=1.5$ $P_z=\epsilon =0$, $\sigma=0.2$, for $\Lambda=0.0001$ and for $\Lambda=0.1$.}
\label{fig:geod3}
\end{figure}
\item{$\sigma=1/4$}

Incoming null particles hit the axis with infinite speed, $\dot\rho\rightarrow\infty$, as $\rho\rightarrow 0$,
    while outgoing particles escape to the singularity $\rho=\pi/\sqrt{3\Lambda}$, attaining constant speed. In particular, this also holds for $V_{1/4}=0$ or $L_z=0$.

\item{$\sigma>1/4$}

Incoming particles approach the axis with infinite speed and negative acceleration. Outgoing particles
move with decreasing speed and negative acceleration, attaining a maximum distance from the axis for
\begin{equation}
P_{max}=\left(\frac{cE}{L_z}\right)^{\Sigma/(4\sigma-1)}, \label{51a}
\end{equation}
from which we can extract $\rho_{max}$. At $\rho_{max}$, the null particle is reflected back to the axis attaining $\dot\rho\rightarrow\infty$.
For $\Lambda=0$, we have from (\ref{51a})
\begin{equation}
\rho_{LCmax}=\left(\frac{cE}{L_z}\right)^{\Sigma/(4\sigma-1)}, \label{51b}
\end{equation}
which is the maximum distance from the $z$ axis attained by the outgoing null particle in the LC spacetime.
From (\ref{51a}) and (\ref{51b}), we have
        \begin{equation}
        \frac{\sqrt{3\Lambda}}{2}\,\rho_{LCmax}=\tan\left(\frac{\sqrt{3\Lambda }}{2}\,\rho_{max}\right), \label{51c}
        \end{equation}
implying  $\rho_{max}<\rho_{LCmax}$, which shows that $\Lambda$ decreases the maximum distance to the axis reached by the null particle (see an example in Figure \ref{fig:geod4}). Again, $\Lambda$ had the opposite effect for $\Lambda<0$.
\end{enumerate}
\begin{figure}[H]
\begin{minipage}{8cm}
\begin{tabular}{c}
\includegraphics[width=8cm]{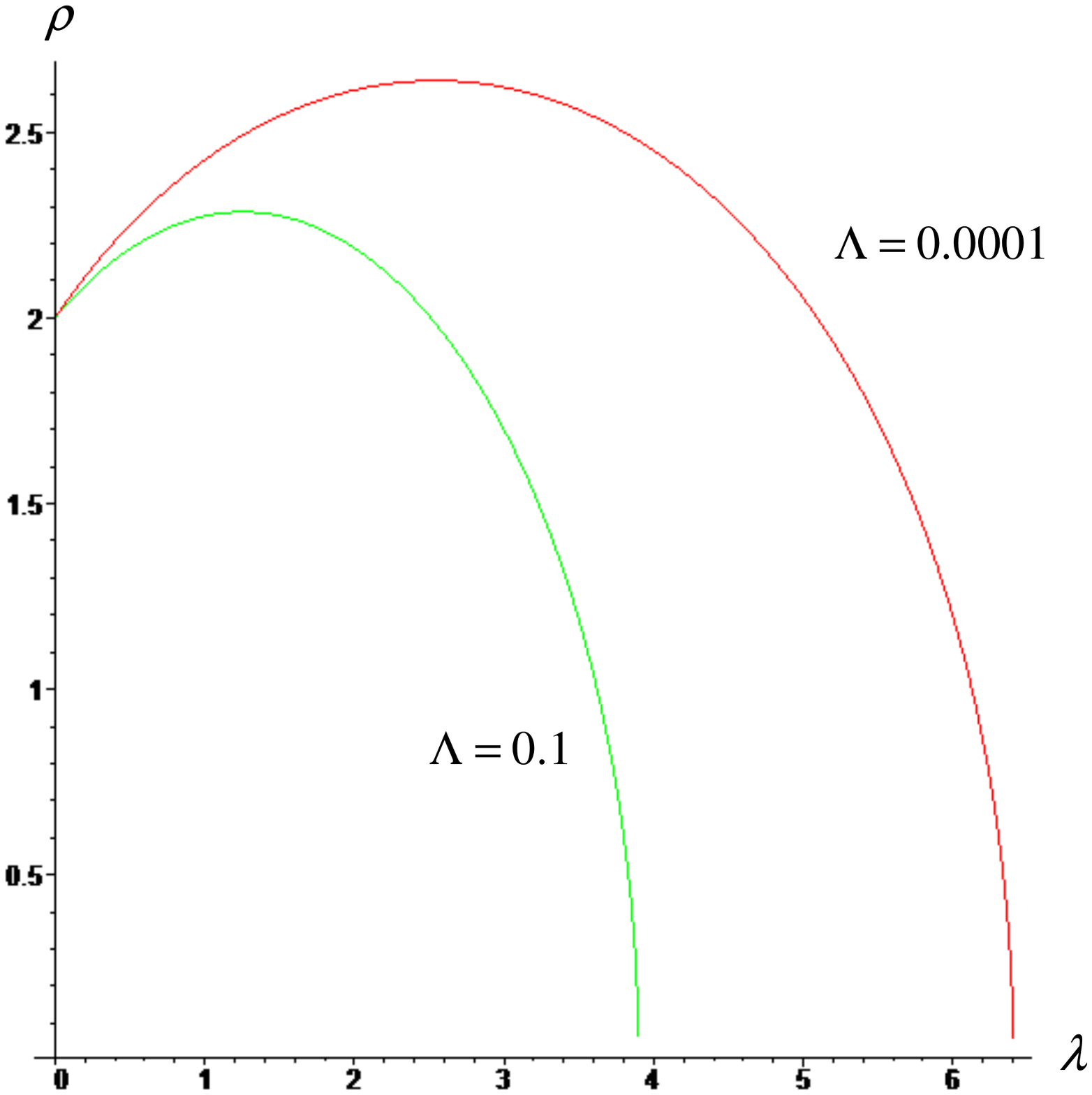} \\
\end{tabular}
\end{minipage}
\begin{minipage}{8cm}
\begin{tabular}{c}
\includegraphics[width=8cm]{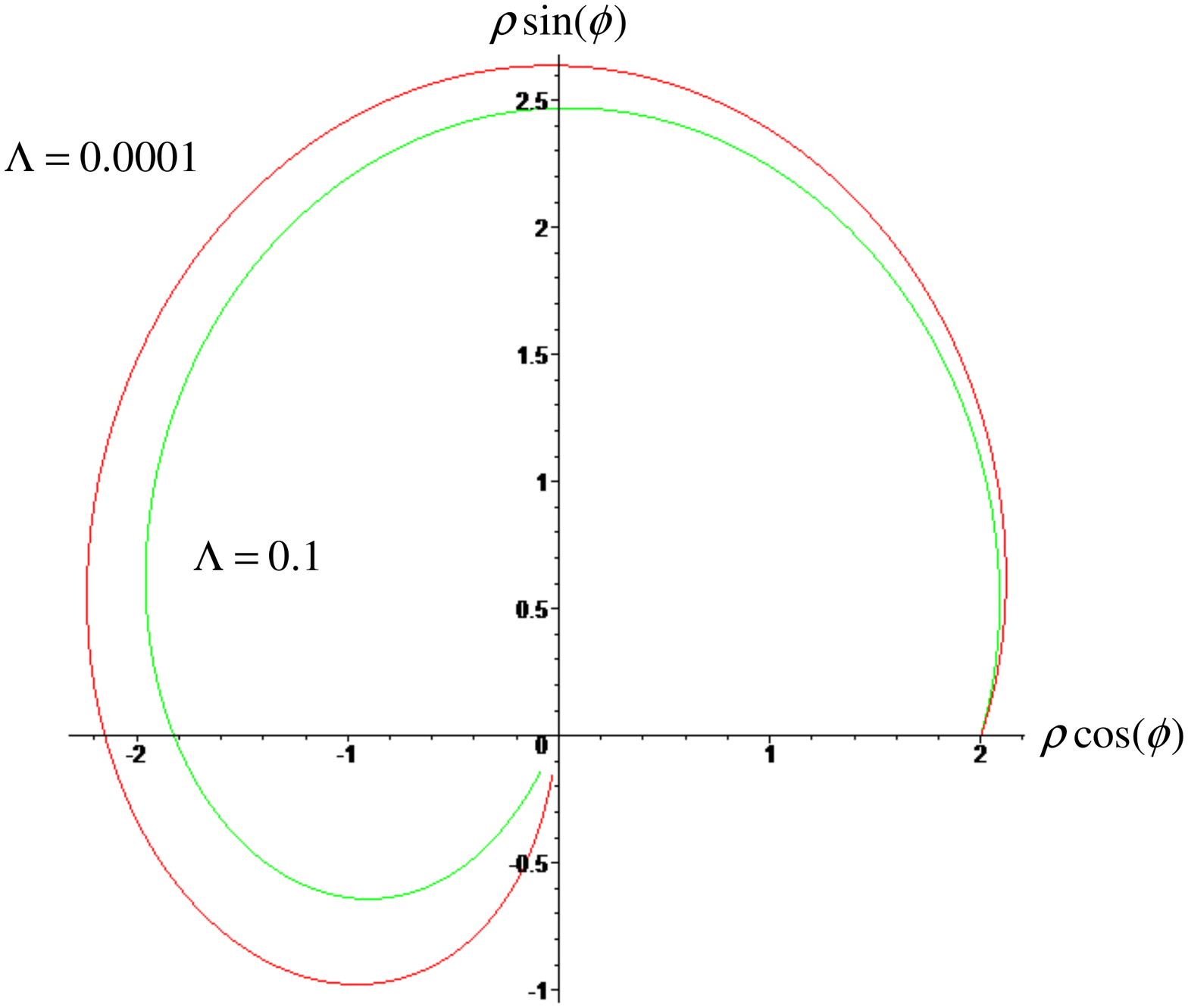}\\
\end{tabular}
\end{minipage}
\caption{ Graphs of the numerical integration of the geodesics' equations along $\rho(\lambda)$, for $E=2$, $L_z=b=c=1$, $P_z=\epsilon =0$, $\sigma=0.4$, for $\Lambda=0.0001$ and for $\Lambda=0.1$.}
\label{fig:geod4}
\end{figure}
\subsubsection{Case $\epsilon\neq 0$}
\label{radial-geo}
In this case, (\ref{24}) is given by
\begin{equation}
V(\rho)=\epsilon Q^{2/3}P^{-2(1-8\sigma+4\sigma^2)/3\Sigma}+\left(\frac{L_z}{c}\right)^2P^{-2(1-4\sigma)/\Sigma} \label{30}\\
\end{equation}
and, taking into account the estimates in Lemma \ref{lemma-estimates}, we split our analysis into the two cases:

\begin{enumerate}
\item{$\sigma<1/4$}

In this case, for $L_z\ne 0$, the geodesics never reach $\rho=0$, since $V(\rho)\to +\infty$ as $\rho\to 0$.

Contrary to the $\Lambda<0$ case, here the equation $V^\star(\rho)=0$ does not always have a solution. This means that $V$ might not have a minimum around which there would be confined orbits. The existence of a solution to  $V^\star(\rho)=0$  depends crucially on $L_z/c$ and $\Lambda$, in the sense that if these quantities are sufficiently large there will be no solution to $V^\star(\rho)=0$. See examples in Figure \ref{fig:geodV}.

When $V^{\star}(\rho)=0$ has two solutions, there will be one such that $V^{\star\star}(\rho)>0$ and another with $V^{\star\star}(\rho)<0$. In those cases, there will always be bounded orbits in the convex part of $V<E^2$, as illustrated in Figure \ref{fig:geod5}.
Increasing $\Lambda$, decreases the value of $\rho=\pi/\sqrt{3\Lambda}$ and, for sufficiently large $\Lambda$, $V^\star=0$ will no longer have two solutions. In that case, bounded geodesics became unbounded and converge to the $\rho=\pi/\sqrt{3\Lambda}$ singularity.
This is a dramatically different behaviour with respect to the $\Lambda<0$ case, where bounded geodesics always remained bounded by changing $\Lambda$.
\begin{figure}[H]
\begin{minipage}{8cm}
\begin{tabular}{c}
\includegraphics[width=8cm]{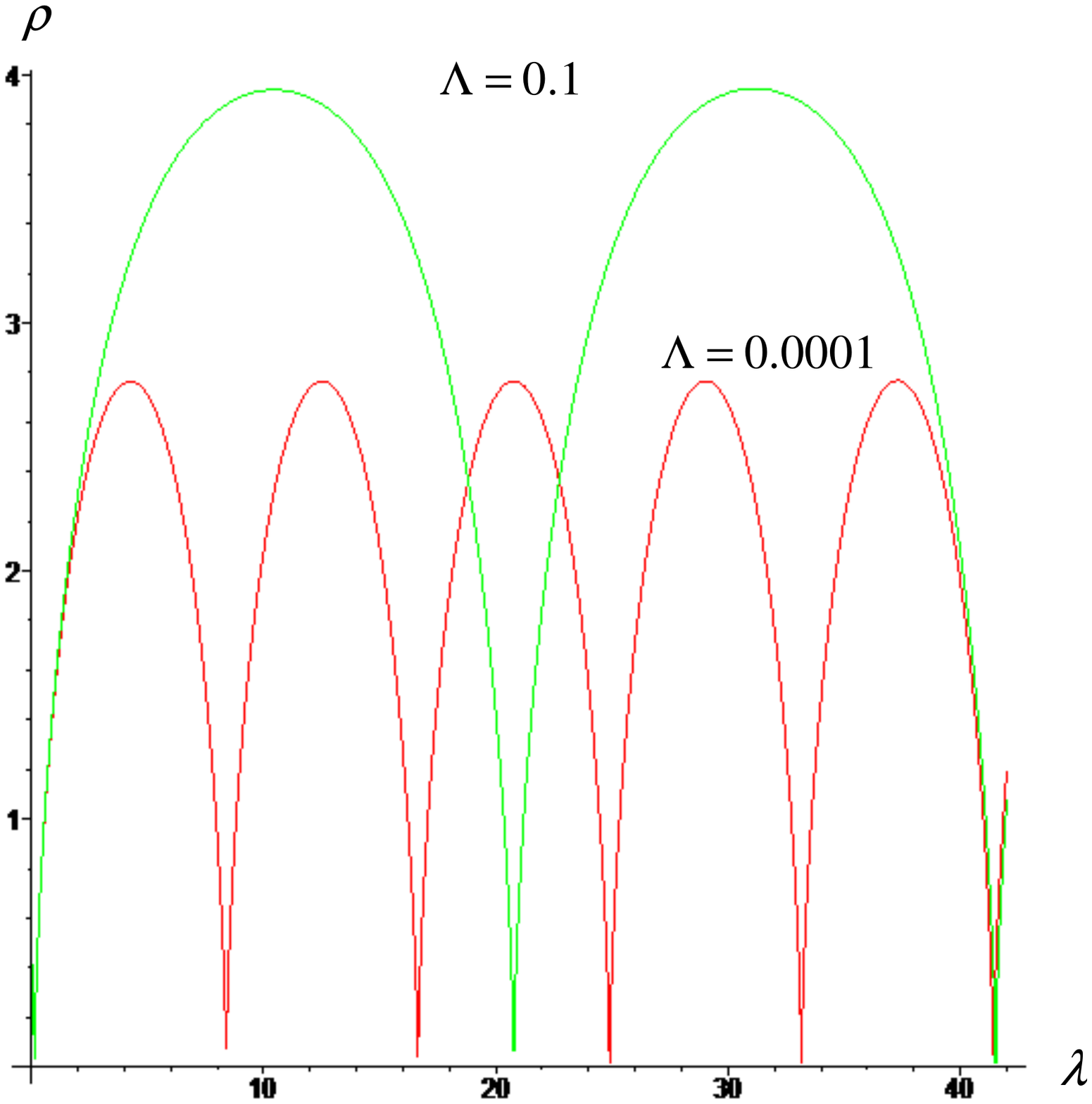} \\
\end{tabular}
\end{minipage}
\begin{minipage}{8cm}
\begin{tabular}{c}
\includegraphics[width=8cm]{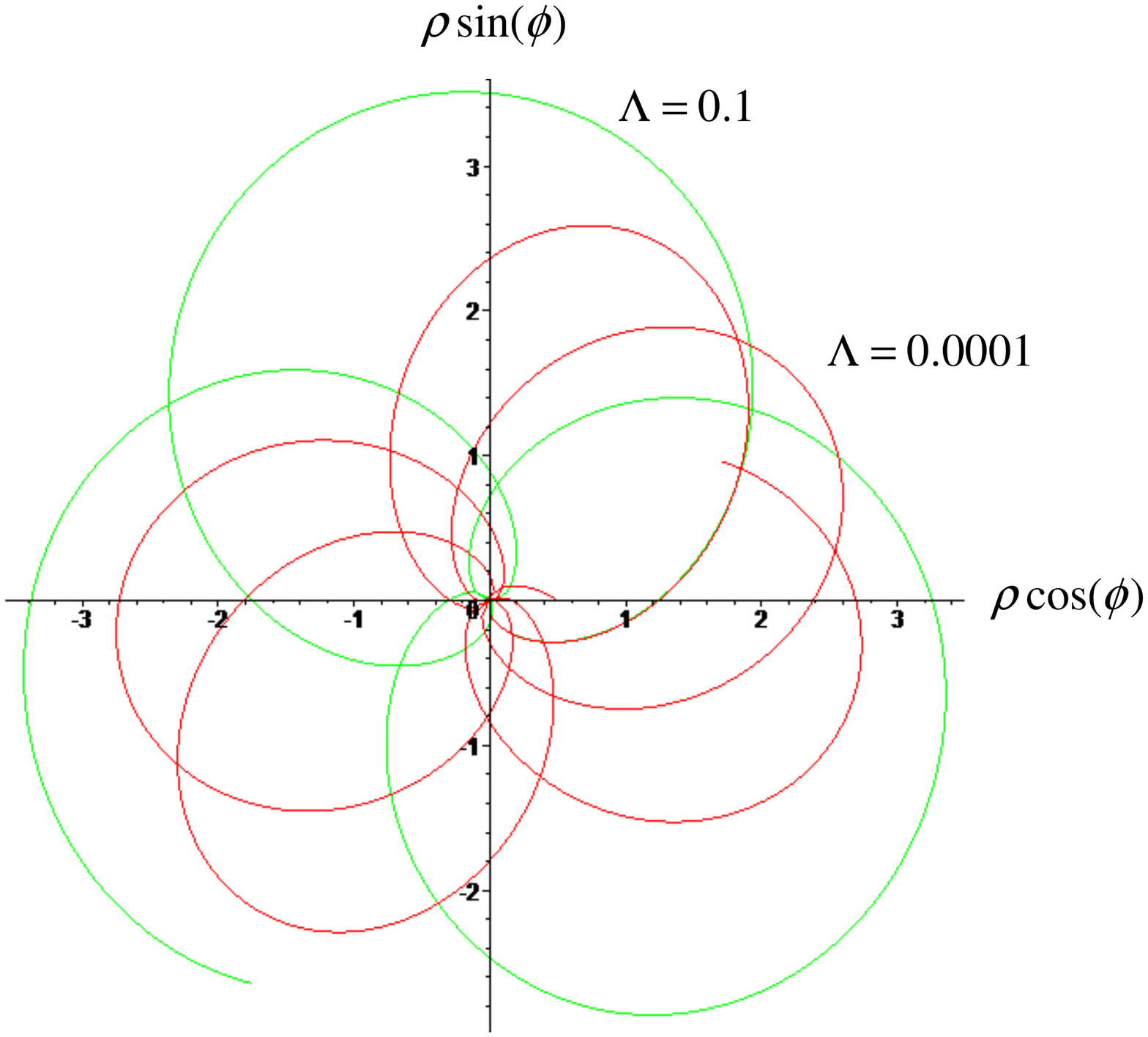}\\
\end{tabular}
\end{minipage}
\caption{ Graphs of the numerical integration of the geodesics' equations along $\rho(\lambda)$, for $E=2$, $L_z=c=\epsilon=1$, $P_z=0$, $\sigma=0.22$, for $\Lambda=0.0001$ and for $\Lambda=0.1$.}
\label{fig:geod5}
\end{figure}

\begin{figure}[H]
\begin{minipage}{8cm}
\begin{tabular}{c}
\includegraphics[width=8cm]{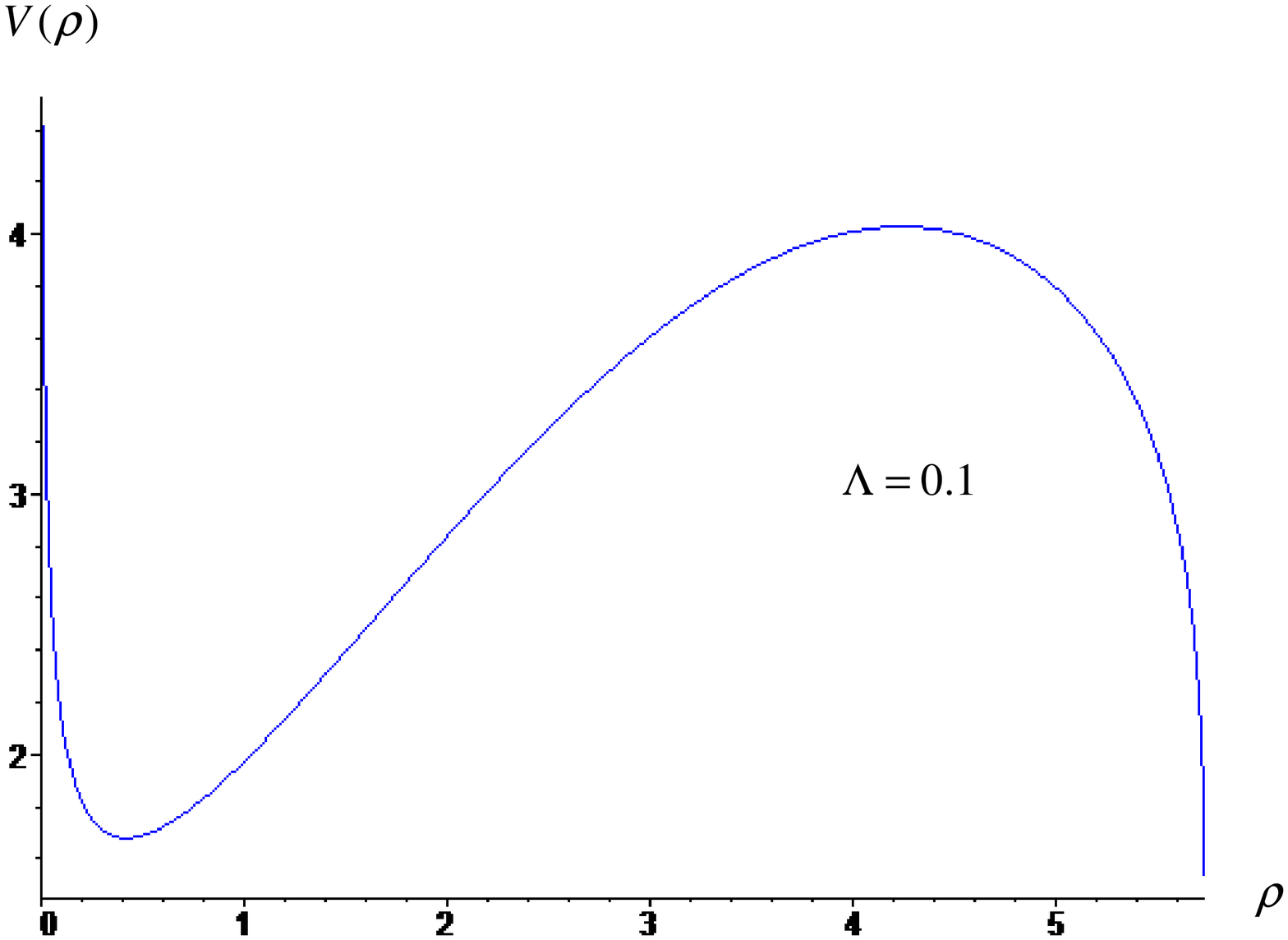}\\
\end{tabular}
\end{minipage}
\begin{minipage}{8cm}
\begin{tabular}{c}
\includegraphics[width=8cm]{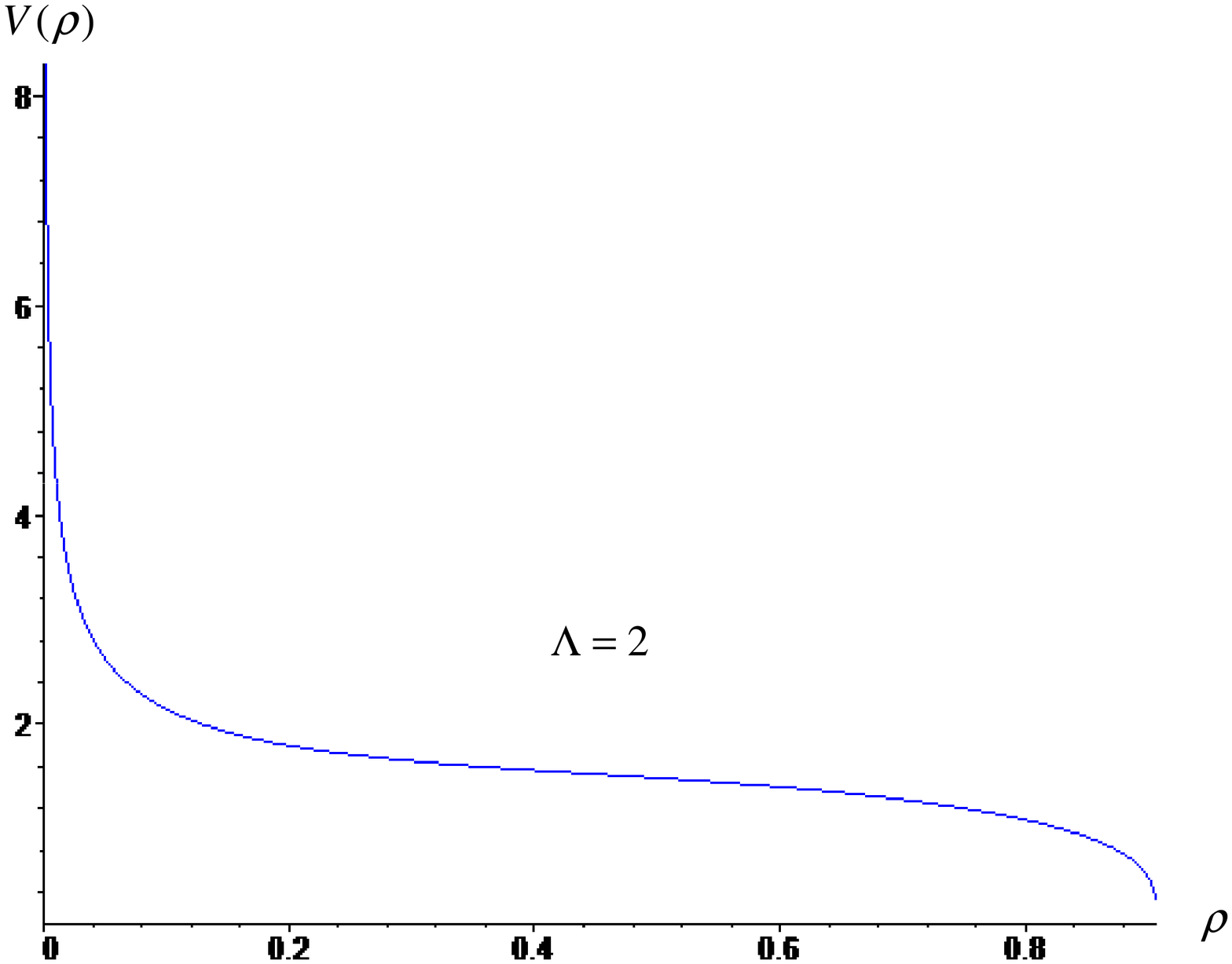}\\
\end{tabular}
\end{minipage}
\caption{ Graph of $V(\rho)$ for $E=2$, $L_z=c=\epsilon=1$, $P_z=0$ and $\sigma=0.22,$ for $\Lambda=0.1$ and for $\Lambda=2$.}
\label{fig:geodV}
\end{figure}

\item{$\sigma\ge1/4$}

In this case, by analysing (\ref{25}) with the estimates (\ref{ineq1}) and (\ref{ineq2}), one can conclude that $V^{\star}$ has no zeros. Considering the limits of $V$ in the approach to the singularities, we find that there is always a maximum value for the radius of outgoing geodesics after which they converge to the singularity $\rho=0$. This is a similar qualitative behaviour as in the $\Lambda<0$ case.

\begin{figure}[H]
\begin{minipage}{8cm}
\begin{tabular}{c}
\includegraphics[width=8cm]{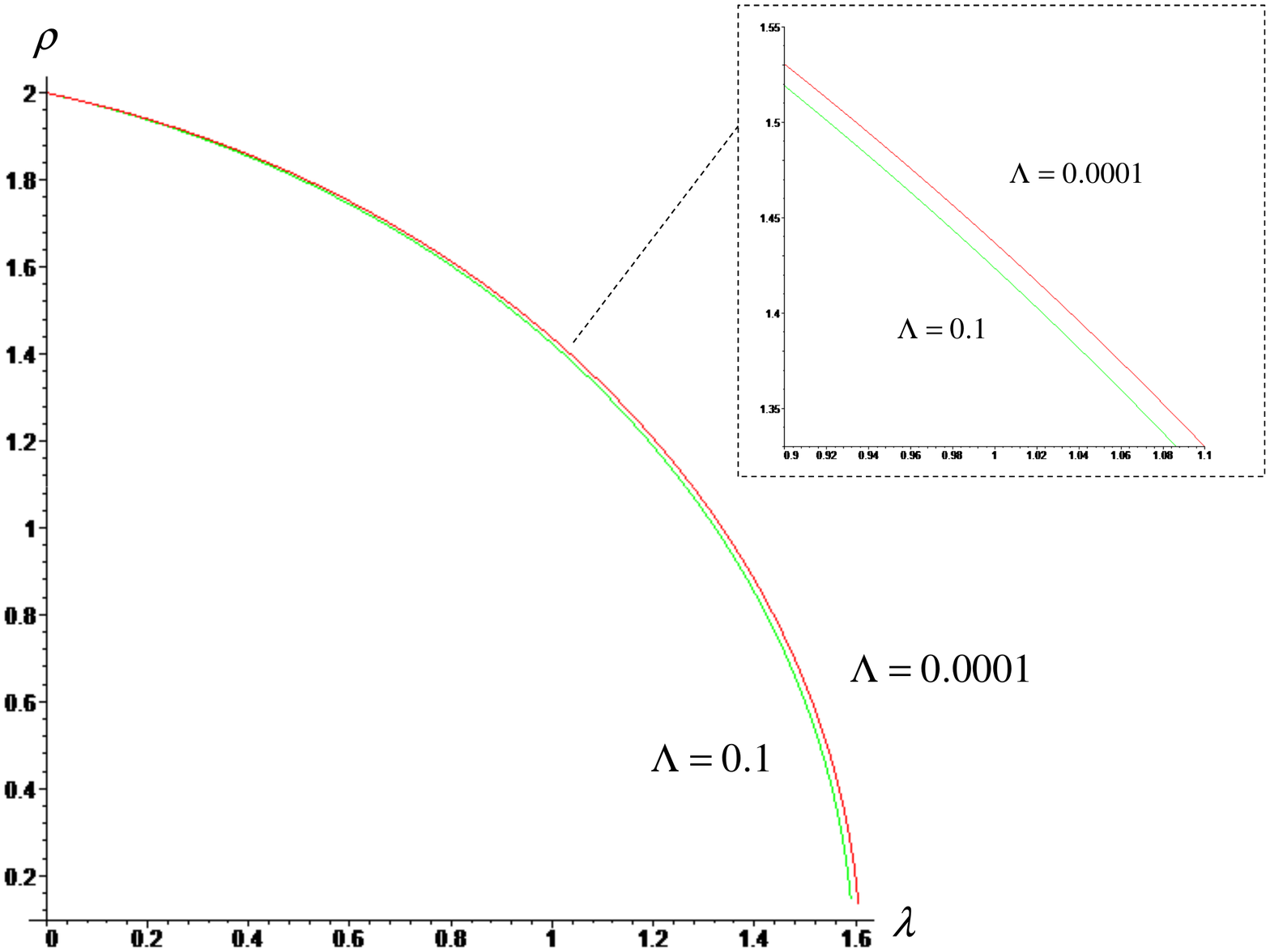} \\
\end{tabular}
\end{minipage}
\begin{minipage}{8cm}
\begin{tabular}{c}
\includegraphics[width=8cm]{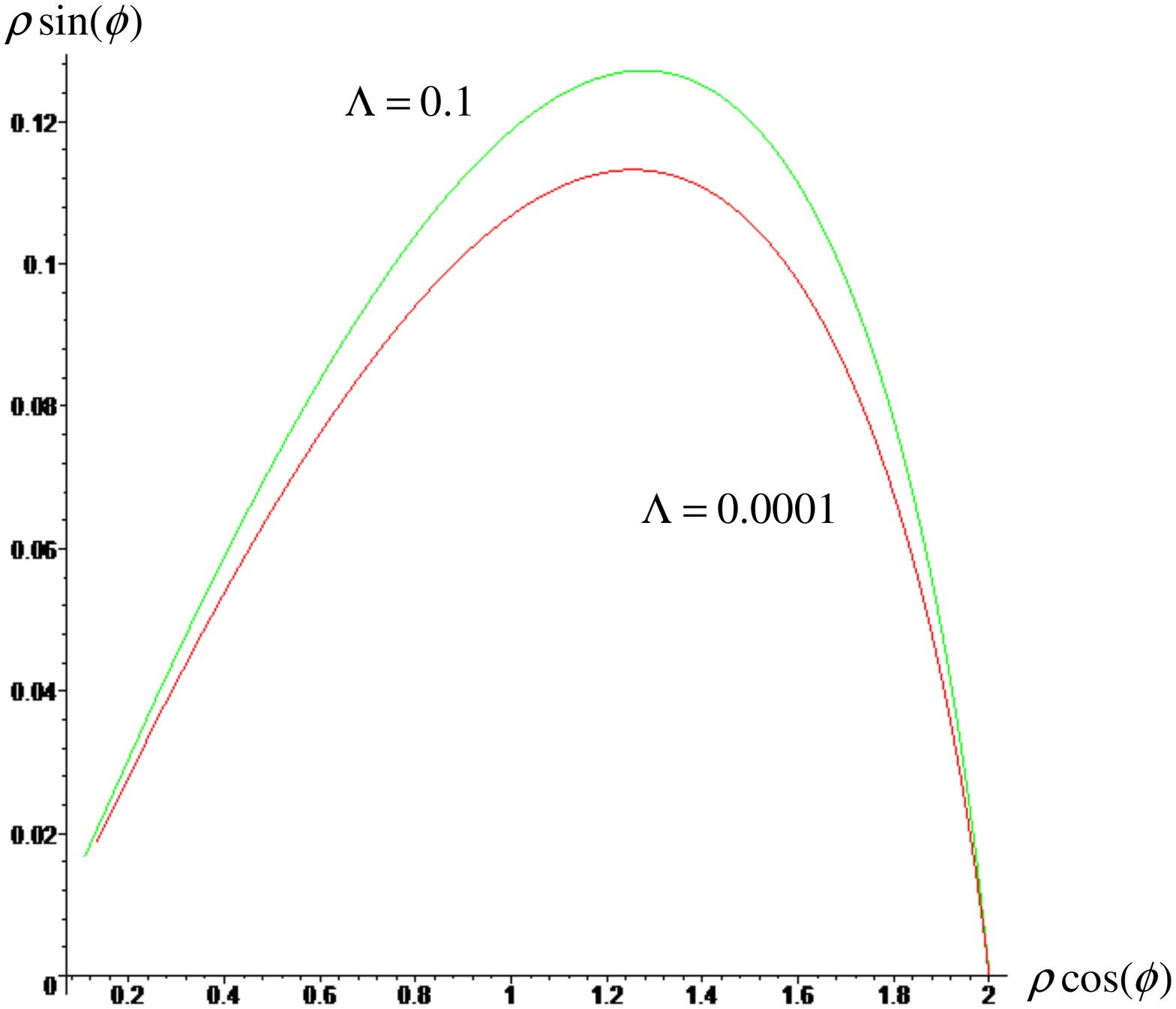}\\
\end{tabular}
\end{minipage}
\caption{ Graphs of the numerical integration of the geodesics' equations along $\rho(\lambda)$, for $E=2$, $L_z=0.1$, $\epsilon=b=c=1$, $P_z=0$, $\sigma=0.4$, for $\Lambda=0.0001$ and for $\Lambda=0.1$.}
\label{fig:geod6}
\end{figure}

\end{enumerate}
Unlike the $\epsilon=0$ case, for $\epsilon\ne 0$ the extreme distances of the geodesics to the $\rho=0$ axis can increase with $\Lambda$, depending on the relative magnitudes of $L_z, c$ and $\sigma$. Examples are depicted in Figures \ref{fig:geod5} and \ref{fig:geod6}.

Taking into account the analysis in \cite{BDMS-2014}, we end this section, highlighting the following result:
\begin{proposition}
Consider geodesics with $P_z=0$. Then, timelike confined non-singular geodesics in LC ($\Lambda= 0$) are (i) stable, for sufficiently small $\Lambda>0$; (ii) unstable, for sufficiently large $\Lambda>0$; (iii) stable, against any change of $\Lambda<0$.

\end{proposition}
\subsection{Non-planar geodesics ($P_z\ne 0$)}
This is the most general case of the geodesics dynamics and this section should be read taking into account the results of Section 4, where the evolution equations for $\dot z$ and $\ddot z$ have been analysed.
\subsubsection{Case $\epsilon=0$}
From (\ref{24}), we have
\begin{equation}
V(\rho)=\left(\frac{P_z}{b}\right)^2 P^{8\sigma(1-\sigma)/\Sigma}+\left(\frac{L_z}{c}\right)^2P^{-2(1-4\sigma)/\Sigma}. \label{35}
\end{equation}

As before, we split our analysis into different intervals for $\sigma$.

\begin{enumerate}
\item{$\sigma<1/4$}

For $L_z\ne 0$, since
\begin{equation}
\label{Lzdiff0}
\underset{\rho\rightarrow 0}{\lim}V(\rho)=\underset{\rho\rightarrow \frac{\pi}{\sqrt{3\Lambda}}}{\lim}V(\rho)=+\infty
\end{equation}
and analysing (\ref{25}) and (\ref{26}), one concludes that $V^{\star}(\rho)=0$ has a solution $\rho=\rho_e$ satisfying $V^{\star\star}(\rho_e)>0$, which indicates that $V(\rho)$ has a minimum.
Consequently, the equation $E^2=V(\rho)$ has two real roots, $\rho_{min}$ and $\rho_{max}$.
 This means that an incoming null particle approaching the $z$ axis is reflected at $\rho=\rho_{min}$, where it attains $\dot\rho=0$, and moves outwards until it attains again $\dot\rho=0$ at $\rho=\rho_{max}$, where it is reflected backwards. This trajectory is repeated endlessly (see an example in Figure \ref{fig:geod7}). For small $\Lambda$, one concludes, after inspecting \eqref{38B}, that increasing $\Lambda$ decreases $\rho_m$. Thus, if we conceive that these particles can constitute something like a jet along $z$, the parameter $\Lambda$ always helps to improve its collimation.

For $L_z=0$, there is only $\rho_{max}$ and no $\rho_{min}$. In that special case, we derive for $\Lambda=0$
\begin{equation}
\rho_{LCmax}=\left(\frac{Eb}{P_z}\right)^{\Sigma/[4\sigma(1-\sigma)]},
\end{equation}
whereas for $\Lambda\ne 0$ we get
\begin{equation}
\frac{\sqrt{3\Lambda}}{2} \rho_{LCmax}=\tan {\left( \frac{\sqrt{3\Lambda}}{2}\rho_{max}\right)},
\end{equation}
which shows how $\Lambda$ decreases $\rho_{max}$. For small $\Lambda$, this can also be seen, explicitly, from \eqref{38B}, which gives
\begin{equation}
\rho_{max}\approx \rho_{LCmax}-\frac{\Lambda}{4}\rho^3_{LCmax}.
\end{equation}

\begin{figure}[H]
\begin{minipage}{8cm}
\begin{tabular}{c}
\includegraphics[width=8cm]{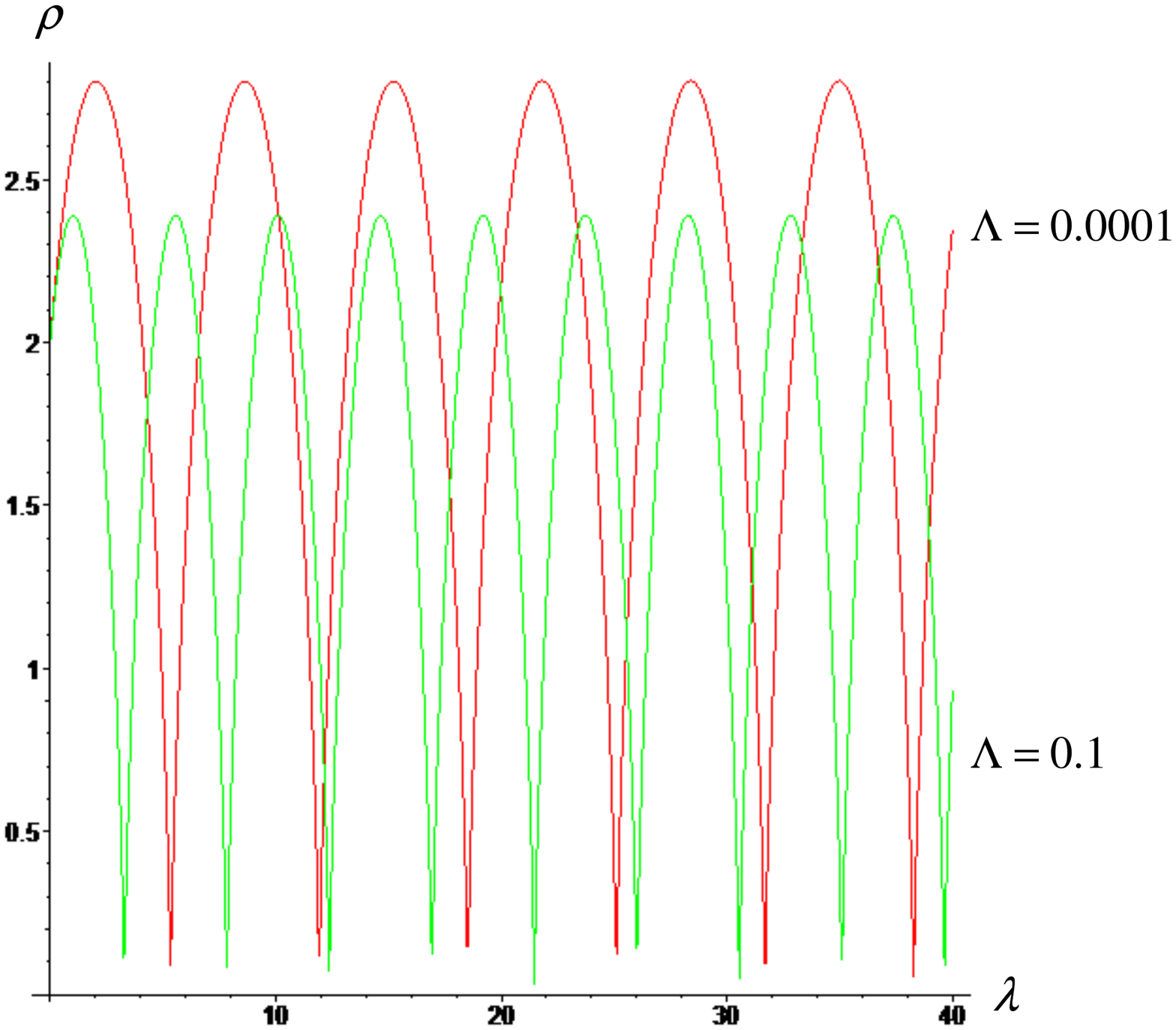} \\
\end{tabular}
\end{minipage}
\begin{minipage}{8cm}
\begin{tabular}{c}
\includegraphics[width=8cm]{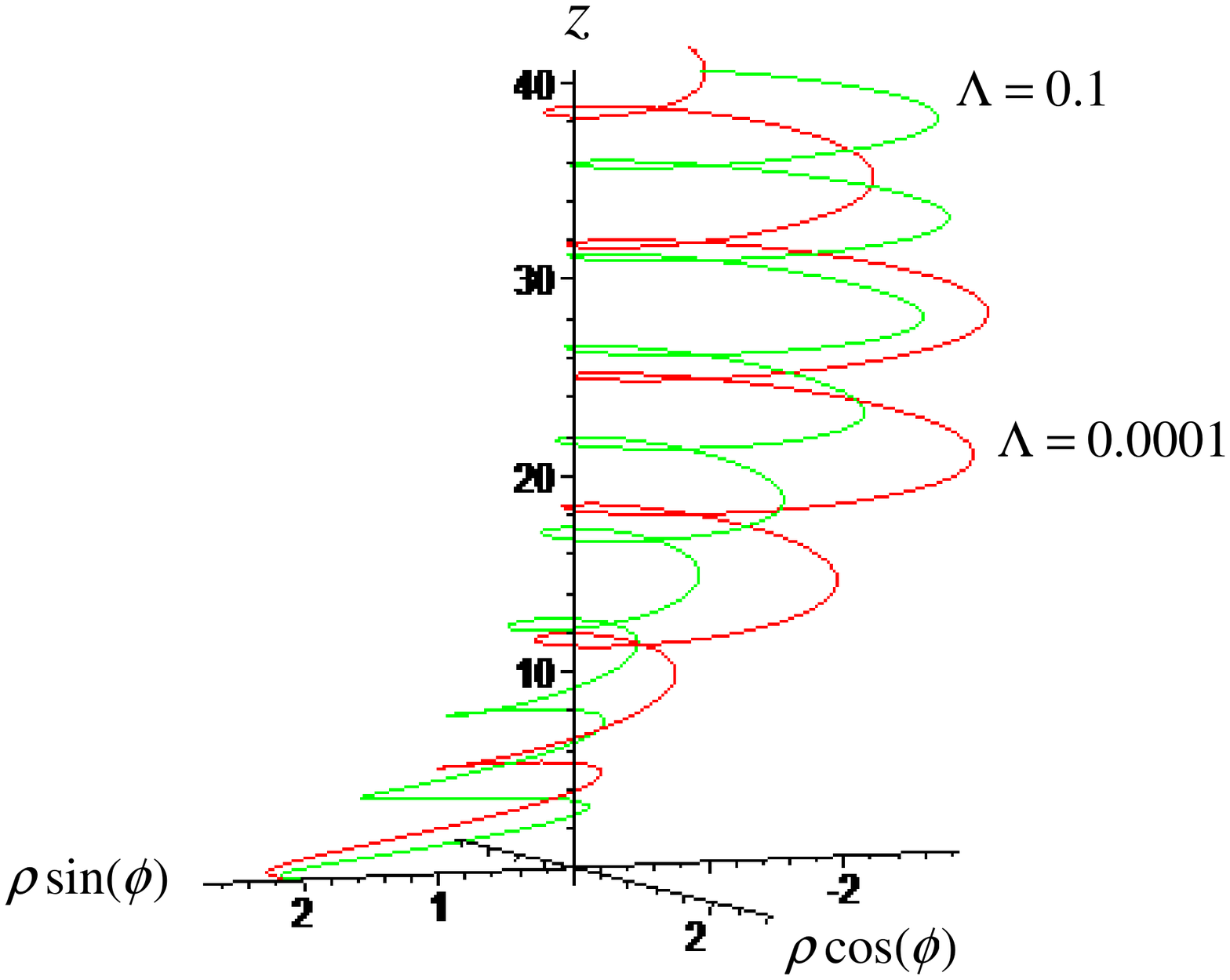}\\
\end{tabular}
\end{minipage}
\caption{ Graphs of the numerical integration of the geodesics' equations along $\rho(\lambda)$, for $E=2$, $L_z=P_z=0.8,$ $b=c=1$, $\epsilon=0$, $\sigma=0.2$, for $\Lambda=0.0001$ and for $\Lambda=0.1$.}
\label{fig:geod7}
\end{figure}
\item{$\sigma\geq 1/4$}

In this case, from Lemma \ref{lemma-estimates} one has
\begin{equation}
\underset{\rho\rightarrow \frac{\pi}{\sqrt{3\Lambda}}}{\lim}V(\rho)=+\infty,
\end{equation}
while the limit of $V$ as $\rho\to 0$ is finite.
Incoming particles approach the axis with infinite speed, $\dot{\rho}\rightarrow +\infty$, and negative acceleration $\ddot{\rho}\rightarrow -\infty$. Outgoing particles move with decreasing speed, $\dot{\rho}\rightarrow 0$, and increasing positive acceleration which reach a maximum where $\dot\rho=0$. Indeed, $\ddot{\rho}\rightarrow +\infty$ as $\rho\rightarrow \pi/\sqrt{3\Lambda}$. Geodesics then reach a maximum radial distance from the axis before moving inwards towards the axis, see Figure \ref{fig:geod8} for an example. For small $\Lambda$ it is easy to see, directly from \eqref{38B}, how $\rho_m$ decreases with increasing $\Lambda$, for any values of $L_z$ and $P_z$.
\begin{figure}[H]
\begin{minipage}{8cm}
\begin{tabular}{c}
\includegraphics[width=8cm]{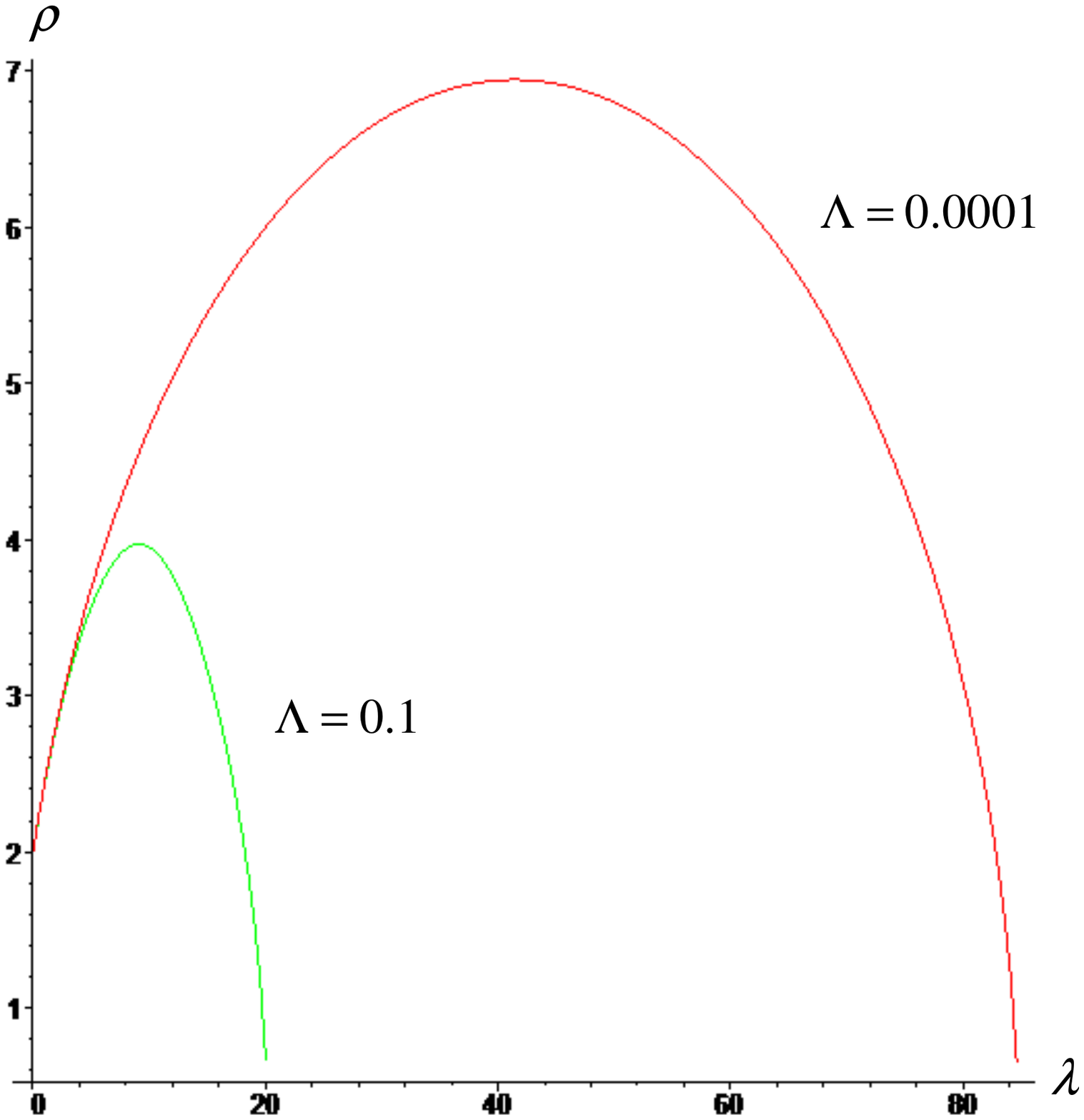} \\
\end{tabular}
\end{minipage}
\begin{minipage}{8cm}
\begin{tabular}{c}
\includegraphics[width=8cm]{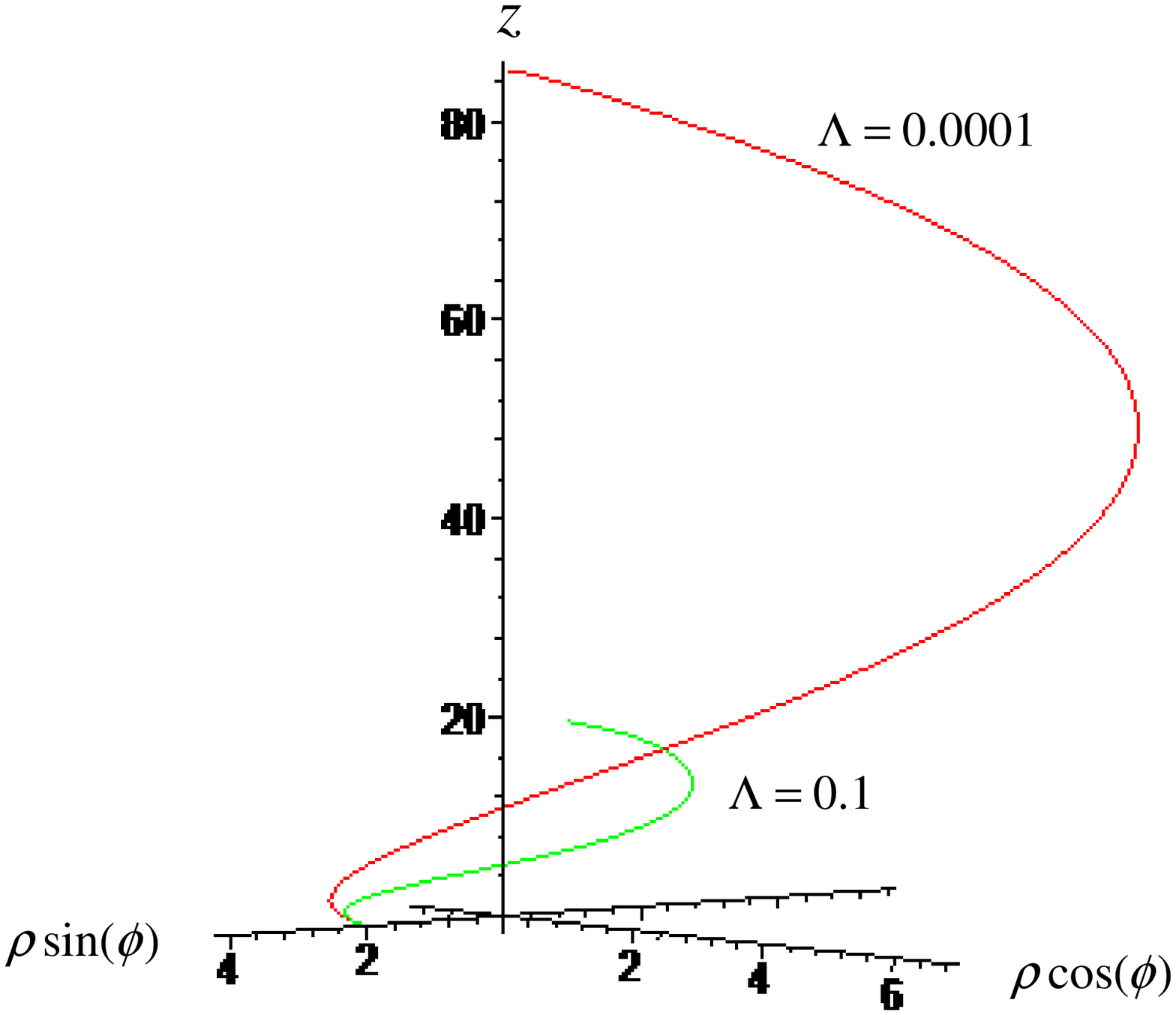}\\
\end{tabular}
\end{minipage}
\caption{ Graphs of the numerical integration of the geodesics' equations along $\rho(\lambda)$, for $E=b=c=1$, $L_z=P_z=0.1$, $\epsilon=0$, $\sigma=0.4$, for $\Lambda=0.0001$ and for $\Lambda=0.1$.}
\label{fig:geod8}
\end{figure}
\end{enumerate}
In the $\Lambda<0$ case, increasing $|\Lambda|$ always increases the extreme radial distances of planar null geodesics to the axis and tends to destabilize their dynamics, while its effect on the non-planar null geodesics' orbits depends on the relative magnitudes of $P_z, L_z, c$ and $\sigma$. On the contrary, for $\Lambda>0$, increasing $\Lambda$ always decreases the extreme distances of the planar and non-planar null geodesics to the axis (see e.g. Figures \ref{fig:geod3}, \ref{fig:geod4}, \ref{fig:geod7} and \ref{fig:geod8}). Again, the parameter $\Lambda$ acts narrowing the beam of null particles along the $z$ direction.

\subsubsection{Case $\epsilon\neq 0$}

\begin{enumerate}

\item{$\sigma<1/4$}

For $L_z\ne 0$, one has
\begin{equation}
\underset{\rho\rightarrow 0}{\lim}V(\rho)=\underset{\rho\rightarrow \frac{\pi}{\sqrt{3\Lambda}}}{\lim}V(\rho)=+\infty
\end{equation}
and this case has some qualitative similarities to the $\epsilon=0$ case. See an example in Figure \ref{fig:geod9}.

\begin{figure}[H]
\begin{minipage}{8cm}
\begin{tabular}{c}
\includegraphics[width=8cm]{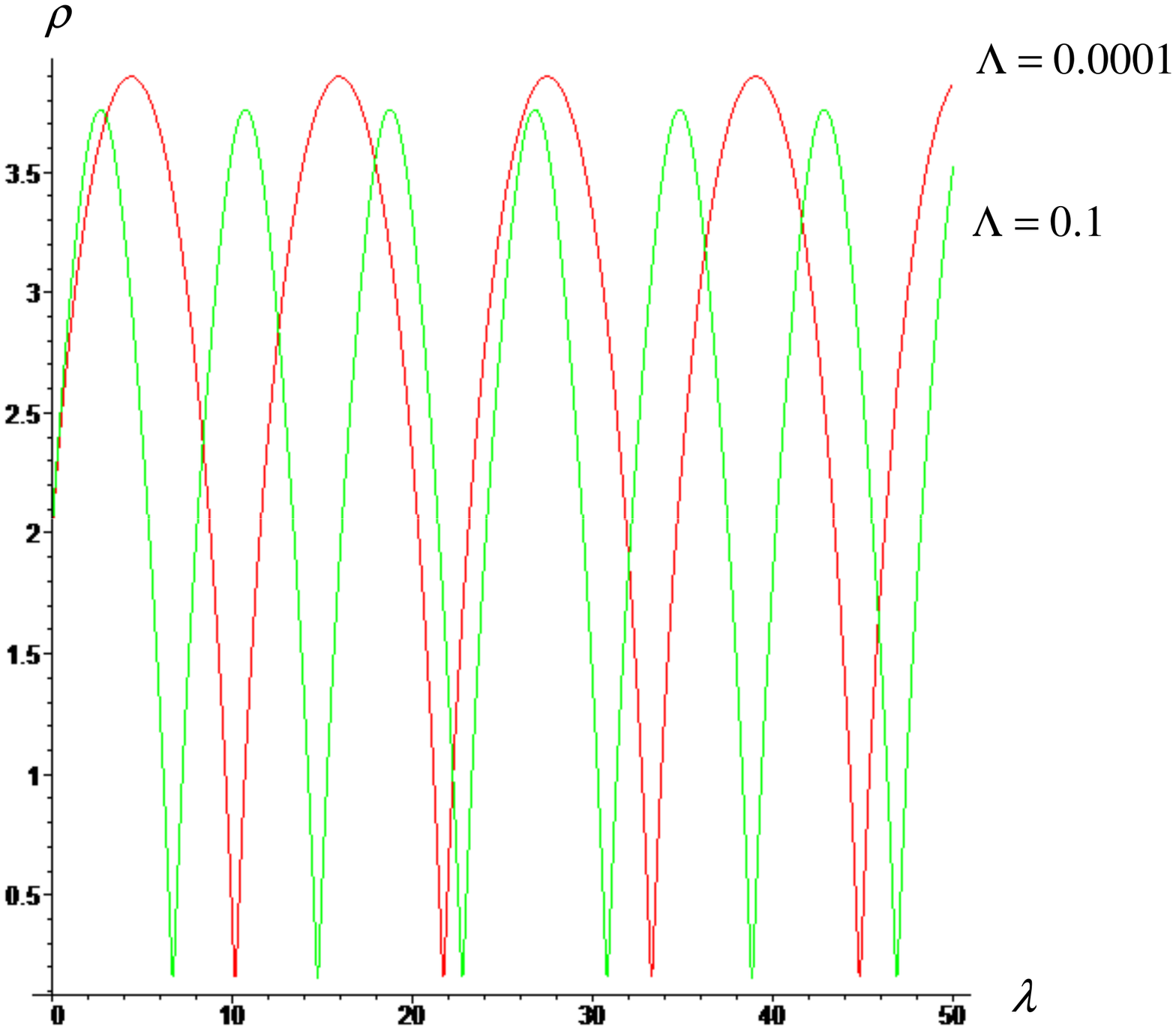} \\
\end{tabular}
\end{minipage}
\begin{minipage}{8cm}
\begin{tabular}{c}
\includegraphics[width=8cm]{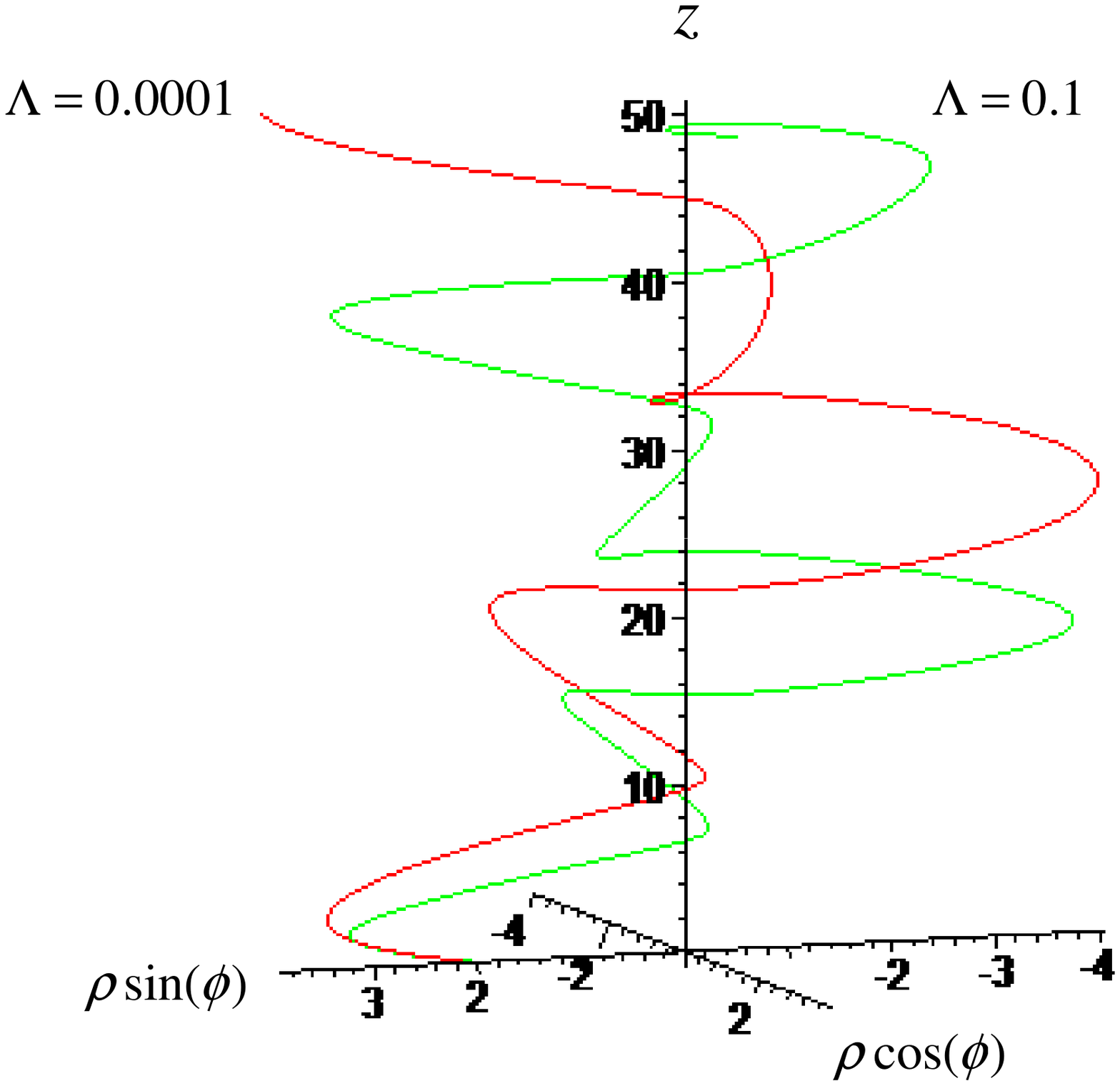}\\
\end{tabular}
\end{minipage}
\caption{ Graphs of the numerical integration of the geodesics' equations along $\rho(\lambda)$, for $E=2$, $\epsilon=b=c=1$, $L_z=0.5, P_z=0.8$, $\sigma=0.1$, for $\Lambda=0.0001$ and for $\Lambda=0.1$.}
\label{fig:geod9}
\end{figure}

Interestingly, one can observe analytically that, depending on the value of $E$, there may exist three regions of confined orbits. An example of this is in Figure \ref{novageod9b}. By comparison with the respective planar case, we conclude that the radial instability present in that case, due to large $\Lambda$, disappears when $P_z\ne 0$ is introduced. In turn, there is now an instability along the $z$ direction, as mentioned in Section 4.

\begin{figure}[H]
\begin{minipage}{5cm}
\begin{tabular}{c}
\includegraphics[width=6cm]{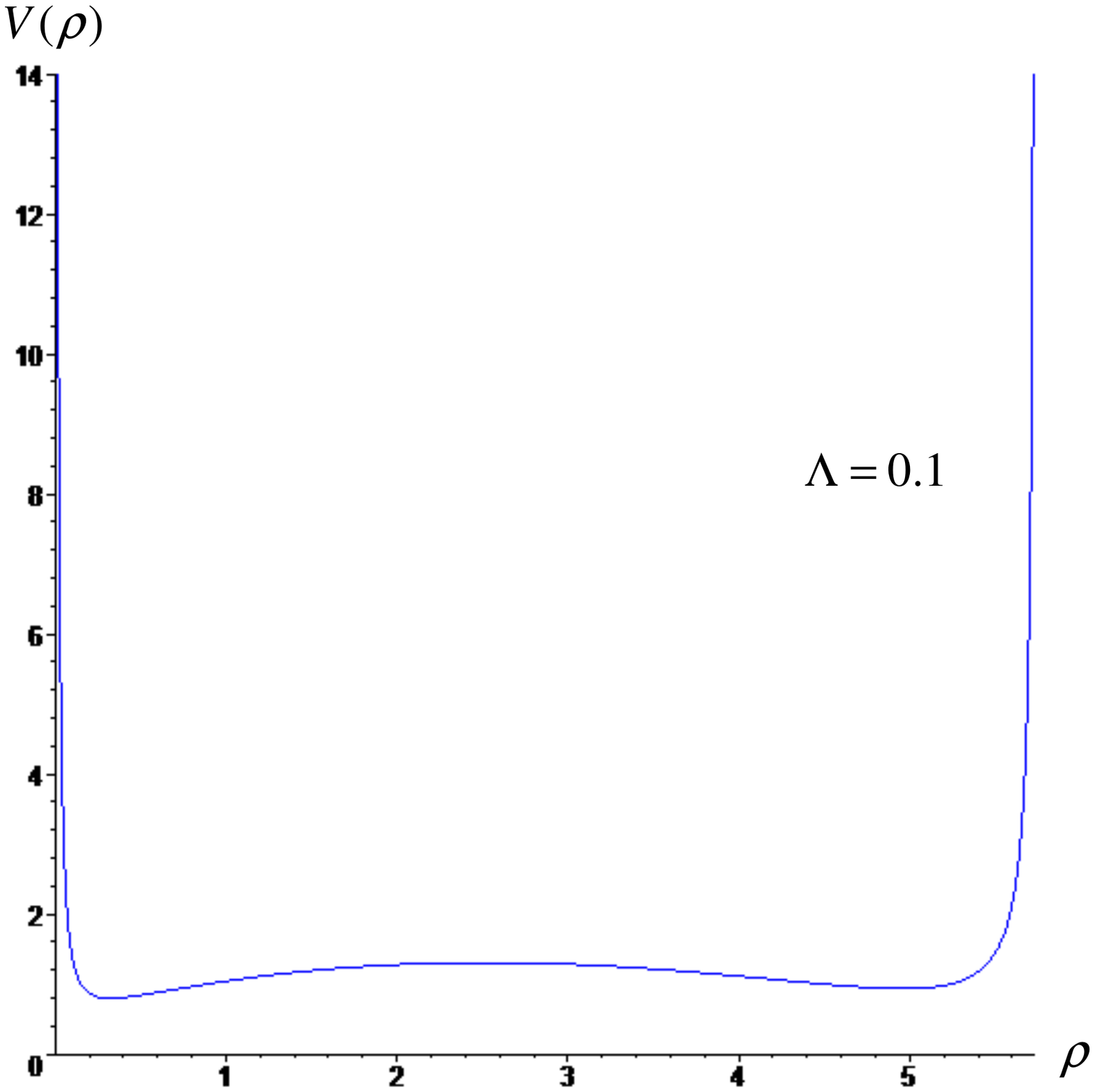} \\
\end{tabular}
\end{minipage}
\begin{minipage}{5cm}
\begin{tabular}{c}
\includegraphics[width=6cm]{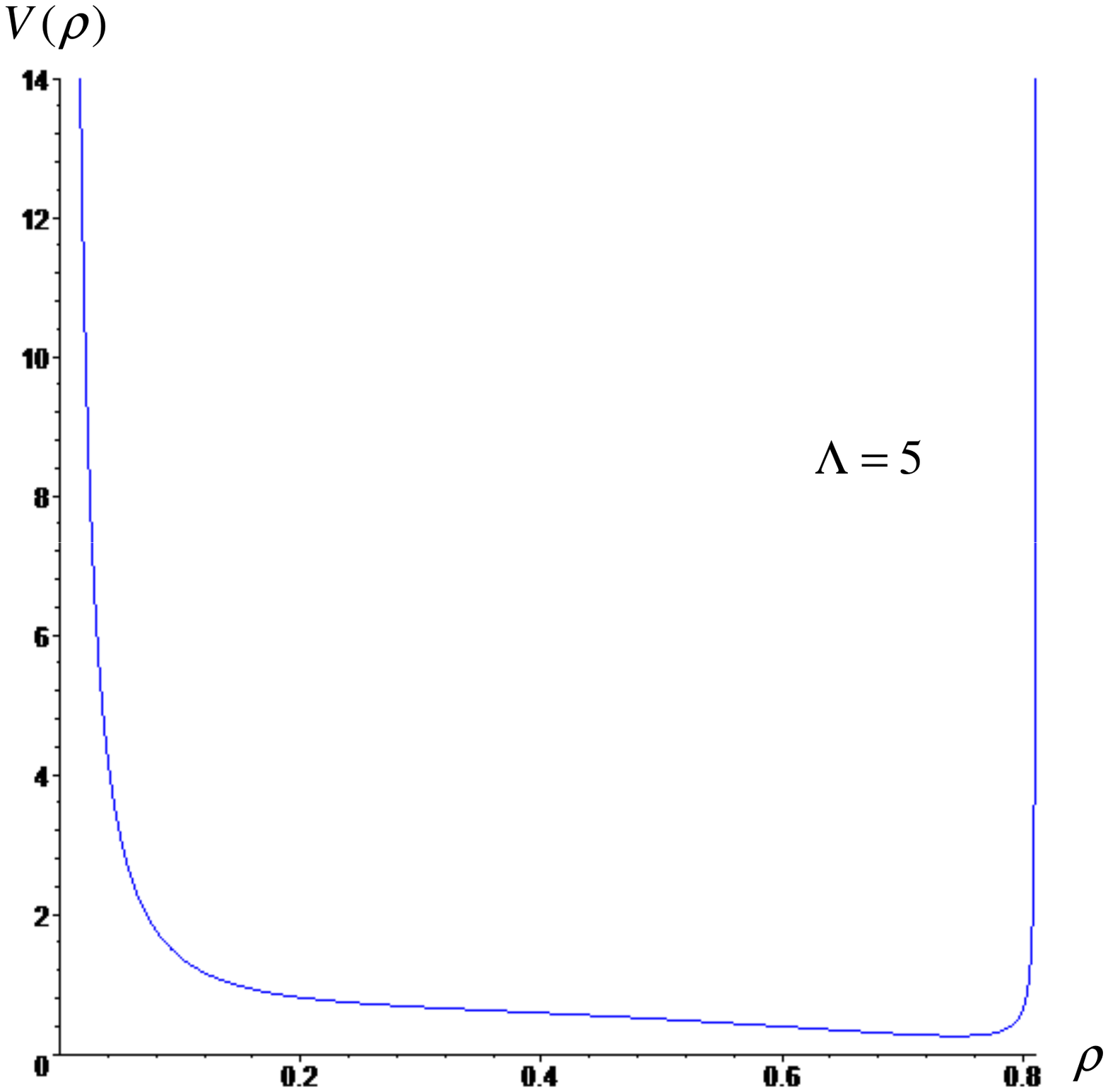}\\
\end{tabular}
\end{minipage}
\begin{minipage}{5cm}
\begin{tabular}{c}
\includegraphics[width=6cm]{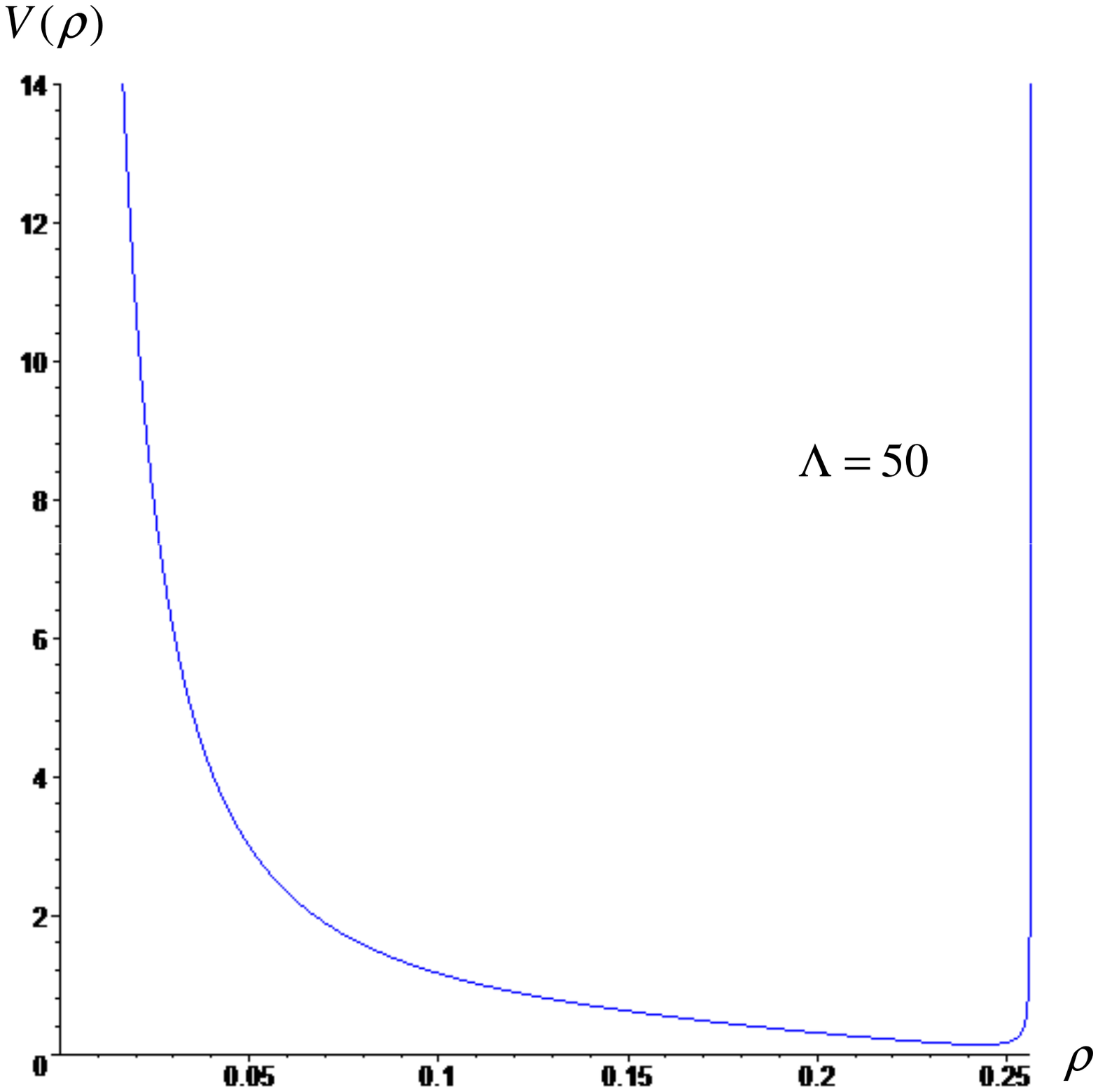}\\
\end{tabular}
\end{minipage}
\caption{ Graph of $V(\rho)$, for $E=2$, $b=c=\epsilon=1$, $P_z=L_z=0.2$, $\sigma=0.1,$ for $\Lambda=0.1,$ $\Lambda=5$ and $\Lambda=50$.}
\label{novageod9b}
\end{figure}

For $L_z=0$, we recall that in the case $\Lambda<0$, there was always orbital confinement between two non-singular $\rho_{min}$ and $\rho_{max}$, which is not the case for $\Lambda>0$ here since, as for $\epsilon=0$, there is $\rho_{max}$, but no $\rho_{min}$.
\item{$\sigma\ge 1/4$}

In this case,
\begin{equation}
\underset{\rho\rightarrow \frac{\pi}{\sqrt{3\Lambda}}}{\lim}V(\rho)=+\infty,
\end{equation}
the limit of $V$ as $\rho\to 0$ is finite, and the dynamics is qualitatively similar to the $\epsilon=0$ case. In particular, $V^\star$ has no zeros in this case and, therefore, there is no geodesic confinement between two non-singular radii. We recall that this is not the case for $\Lambda<0$ where the geodesics confinement exists for $\sigma>1/4$ (see \cite{BDMS-2014}).

\begin{figure}[H]
\begin{minipage}{8cm}
\begin{tabular}{c}
\includegraphics[width=8cm]{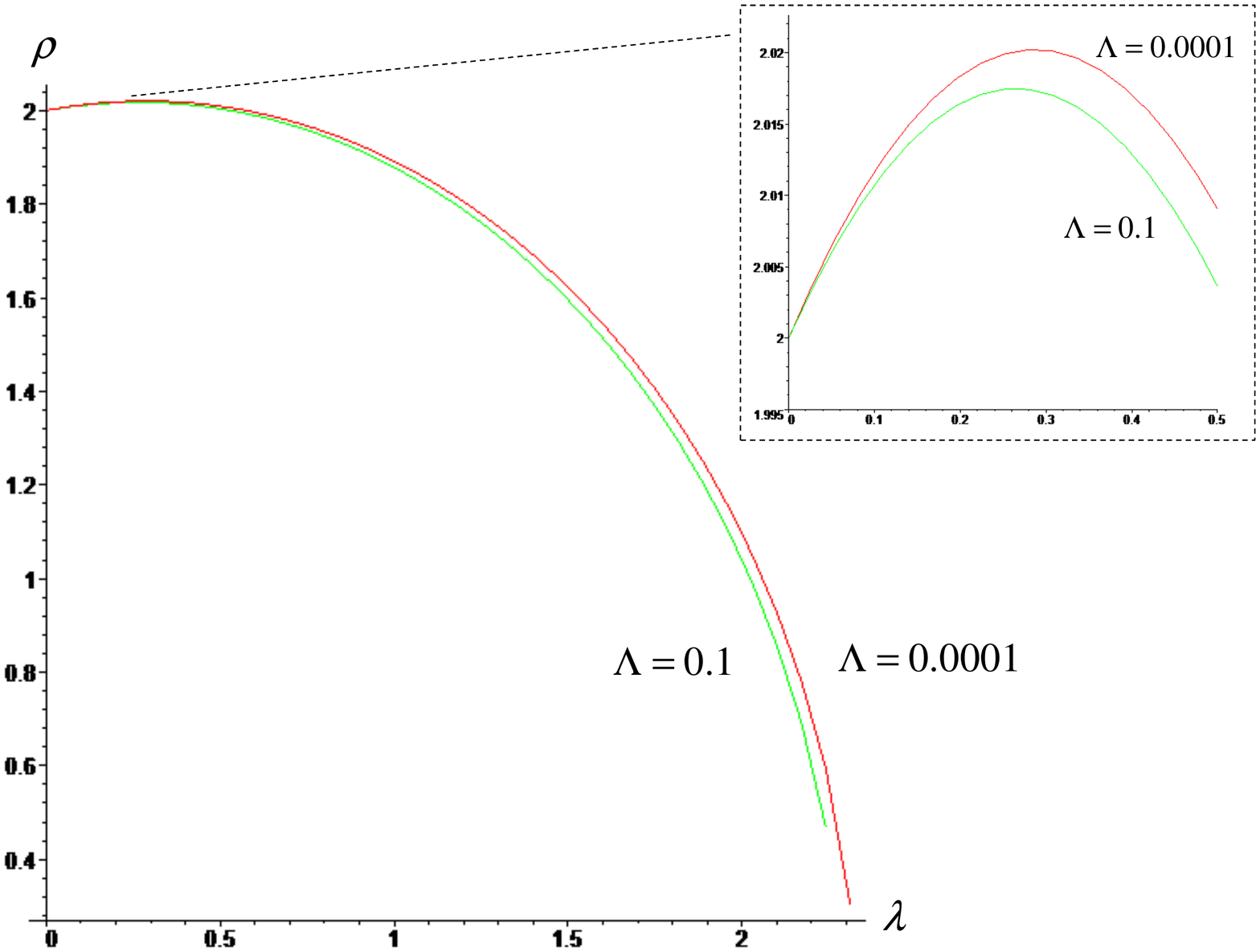} \\
\end{tabular}
\end{minipage}
\begin{minipage}{8cm}
\begin{tabular}{c}
\includegraphics[width=8cm]{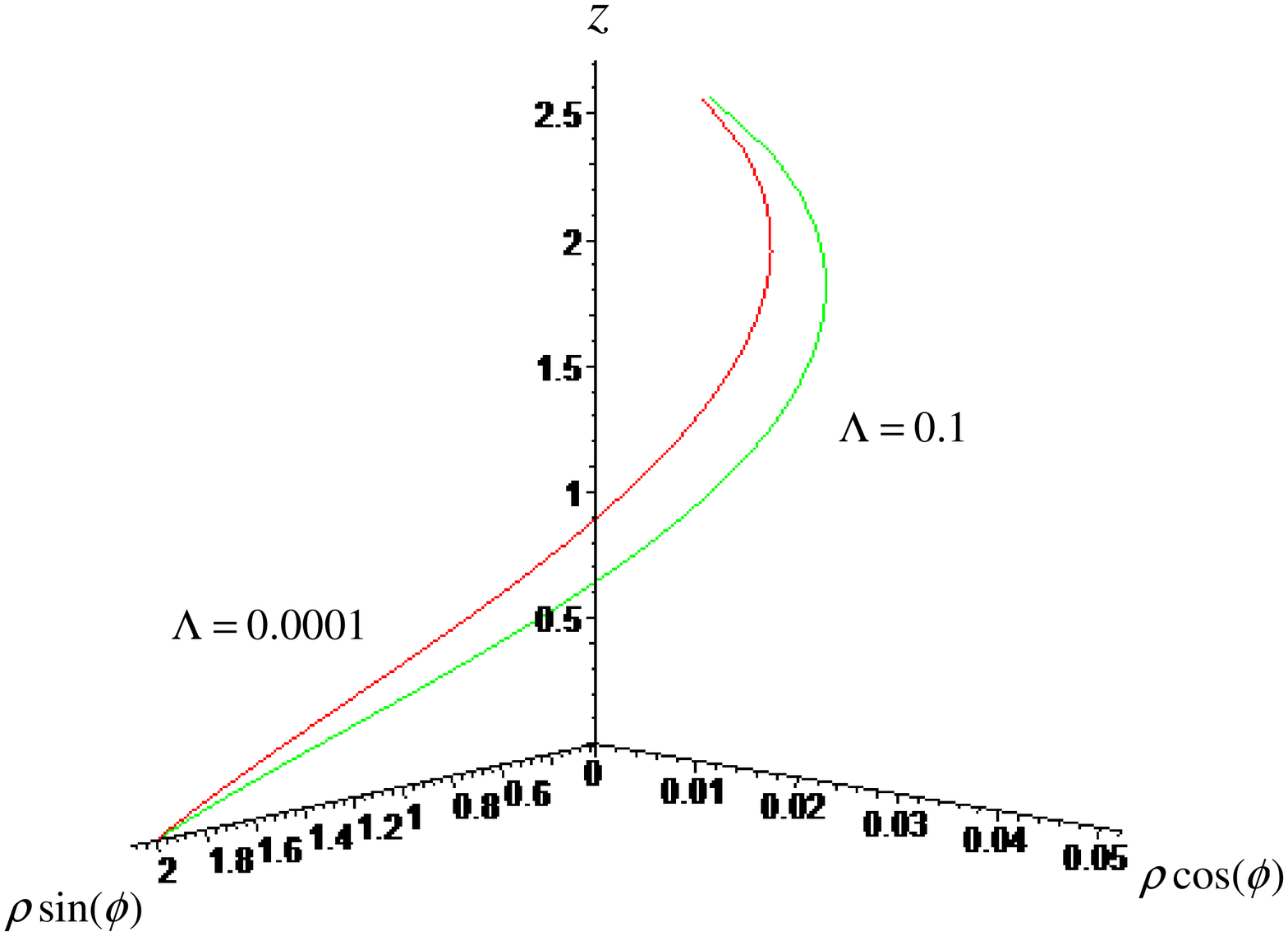}\\
\end{tabular}
\end{minipage}
\caption{ Graphs of the numerical integration of the geodesics' equations along $\rho(\lambda)$, for $E=2$, $\epsilon=b=c=1$, $L_z=0.25,$ $P_z=0.05$, $\sigma=0.4$, for $\Lambda=0.0001$ and for $\Lambda=0.1$.}
\label{fig:geod10}
\end{figure}
\end{enumerate}

An interesting difference between planar and non-planar null geodesics is that, while in the former case $\Lambda$ always decreases the extreme distances of the geodesics to the axis and tends to destabilise their dynamics, in the latter case the effect of $\Lambda$ on the geodesics' orbits, for $\epsilon\ne 0$, depends on the relative magnitudes of $P_z, L_z, c$ and $\sigma$. This can be seen through (\ref{38B}), as linear order effect, and in the non-linear examples of Figures \ref{fig:geod9} and \ref{fig:geod10}.

The effect of $\Lambda$ on the orbits is harder to derive explicitly for the case $\epsilon\ne 0$, but it becomes clear in some special subcases. For instance, for small $\Lambda$, in the $\sigma=0$ case
\begin{eqnarray}
\rho_{m}\approx\rho_{LCm} -\;\frac{\Lambda}{4}\; \left[\left(\frac{L_z}{c}\right)^2
+ \epsilon  \rho_{LCm}^2 \right] \left(\frac{c}{L_z}\right)^2\rho_{LCm}^3
\end{eqnarray}
and for $\sigma=1/2$
\begin{equation}
\rho_{m}\approx\rho_{LCm} -\frac{\Lambda}{4} \left[ \left(\frac{L_z}{c}\right)^2+\left(\frac{P_z}{b}\right)^2\right] \left[ \left(\frac{L_z}{c}\right)^2+\epsilon+\left(\frac{P_z}{b}\right)^2\right]^{-1} \rho_{LCm}^3,
\end{equation}
which shows, in those cases, and at linear order, how $\Lambda$ decreases the extreme distances $\rho_m$.

Taking into account the analysis in \cite{BDMS-2014} and the fact that non-planar geodesics for $\Lambda >0$ are (i) radially confined and non-singular, for $\sigma<1/4$ and $L_z\ne 0$; (ii) never confined, if  $\sigma\ge 1/4$ or $L_z=0$, we deduce:
\begin{proposition}
Consider geodesics with $P_z\ne 0$. Then, radially confined non-singular geodesics in LC ($\Lambda=0$) are
(i) Stable against the introduction of any $\Lambda>0$ (ii) Stable against any $\Lambda<0$ in the case of timelike geodesics, while null geodesics become unstable for $\Lambda<0$ and large enough $E$.
\end{proposition}

\section{Conclusion}

In this paper, we have studied the LT spacetime, which is the $\Lambda$-vacuum solution of the EFEs in cylindrical symmetry. Since it is the counterpart of the Schwarzschild solution in a non-spherically symmetric setting, the LT has a number of physically interesting applications, which we enumerate in the introduction.

When $\Lambda>0$, the solution contains two singularities and this is a fundamental distinguishing feature from the $\Lambda=0$ and $\Lambda<0$ cases analysed before \cite{Banerjee, BDMS-2014}. In order to construct a globally non-singular spacetime we have cut the $\Lambda>0$ LT metric at its two singular ends and glued it to an interior anisotropic fluid source, in one end, and to an exterior $\Lambda$-dust Einstein universe, on the other.

We have then investigated the geodesic motion on a LT spacetime with $\Lambda>0$. To do that, we have derived asymptotic estimates for an effective potential and radial velocity functions which enabled a qualitative analysis of the geodesics system of equations.

We studied the effect of $\Lambda$ on the geodesic orbital stability by considering the effect of $\Lambda$ on bounded orbits.
We have found that $\Lambda=0$ planar timelike confined geodesics are unstable against the introduction of a sufficiently large $\Lambda$, in the sense that the radialy bounded orbits become  radialy unbounded. This instability created by $\Lambda$, therefore, turns bounded geodesics in LC into unbounded geodesics in LT which can, in the latter case, be extended through the outer Einstein universe and reach observers located further away from the source.
However, those geodesic orbits are stable for small enough $\Lambda$. Furthermore, any $\Lambda=0$ non-planar radially confined geodesics also remain radially confined after the introduction of any positive $\Lambda$.

Nevertheless, for any real value of $\Lambda$, if a non-planar geodesic does not reach the central singularity in finite time, it always becomes unbounded along the axial direction.
In particular, combining the results of Lemma 2 and the conclusions following equations (\ref{rhomrholcm}) and (\ref{Lzdiff0}), we may state that, for $\sigma  <1/4$, the LT solution may represent a cylindrical layer of $\Lambda$-vacuum, trapping geodesic beams along the $z$-direction, whose thickness is decreasing with $\Lambda$, at least for small values of $\Lambda$. Therefore, as mentioned in the introduction, we can think of the global solution studied here
as an approximate model for a region where there is a jet, e.g. an extragalactic jet, faraway from the galactic disc and perpendicular to it. 
For $\sigma \geq 1/4$, the parameter $\Lambda$ has a similar effect concerning the maximum radius of the cylindrical layer, but the geodesics are reflected toward to the axis, reaching it in finite time.


Based on our results, we can then speculate that the collimation of some material jets 
can be reinforced by purely gravitational effects due the presence of a small positive cosmological constant. This is in agreement with the results of \cite{collimation}.

For the circular geodesics presented in Figure 4, we can consider
$\Lambda=0.1pc^{-2}$, $\Lambda=10^{-4}pc^{-2}$ and, more
realistically, $\Lambda=10^{-19}pc^{-2}$. This gives proper radius
of about $1.1pc$, $1.2885pc$ and $1.2887pc$, respectively, which would be within 
realistic values for some extragalactic jets (see e.g. \cite{Asada, sch}).


\section*{Acknowledgments}
IB and FM thank CMAT, Univ. Minho, for support through the FEDER Funds COMPETE and FCT Projects Est-C/MAT/UI0013/2011 and Est-OE/MAT/UI0013/2014. FM is also supported by FCT project CERN/FP/123609/2011. FM and NOS thank the warm hospitality from Instituto de F\'isica, UERJ, Rio de Janeiro, where this work was completed. MFAdaSilva acknowledges the financial support from FAPERJ (no. E-26/171.754/2000, E-26/171.533.2002, E-26/170.951/2006, E-26/110.432/2009 and E-26/111.714/2010), Conselho Nacional de Desenvolvimento Cient\'{i}fico e Tecnol\'ogico - CNPq - Brazil (no. 450572/2009-9, 301973/2009-1 and 477268/2010-2) and Financiadora de Estudos e Projetos - FINEP - Brazil.

\end{document}